\documentclass[%
reprint,
amsmath,amssymb,
aps,
pra,
]{revtex4-2}
\usepackage{graphicx}
\usepackage{dcolumn}
\usepackage{bm}
\usepackage{hyperref}
\usepackage[mathlines]{lineno}
\usepackage{booktabs}

\usepackage{xcolor}
\usepackage[normalem]{ulem}

\begin{document}
	
	\title{Unconventional quantum spin mixing of two spin-1 atoms \\beyond the single-mode approximation}
	
	\author{Yanhong Liu$^{1, 2, 3}$}
	\author{Fulin Deng$^{4}$}
	\author{Haoxue Qiao$^{1}$}
	\email{Email: qhx@whu.edu.cn}
	\author{Wenxian Zhang$^{2}$}
	\email{Email: wxzhang@hznu.edu.cn}
	
	\affiliation{$^1$School of Physics and Technology, Wuhan University, Wuhan, Hubei 430072, China\\$^2$School of Physics,  Hangzhou Normal University, Hangzhou, Zhejiang 311121, China\\$^3$Hefei National Laboratory, Hefei, Anhui 230088, China\\$^4$Institute of Theoretical Physics, Chinese Academy of Sciences, Beijing 100190, China}
	\date{\today}
	
	\begin{abstract}
		The single-mode approximation (SMA) has been widely employed in ultracold spinor gases, where all spin components share the same spatial mode. By adopting generalized Van Vleck (GVV) perturbation theory, we achieve a relative error of $10^{-6}$ in the ground-state energy with a reduced effective Hilbert space dimension of $\sim 10^3$ for two spin-1 bosons confined in a one-dimensional hard-wall trap. With the accurate GVV method, we explore the validity of the SMA across ground-state properties and quench spin dynamics for two $^{7}\text{Li}$, $^{87}\text{Rb}$, or $^{23}\text{Na}$ atoms. Numerical results show that the quadratic Zeeman energy $q$ provides a powerful control knob: by increasing $|q|$, the fidelity between the spin components in the ground state gradually decreases, leading eventually to the breakdown of the SMA (due to the competition between the kinetic energy and the quadratic Zeeman energy). Quench spin dynamics with abruptly increased $q$ further reveals multiple resonances, where the SMA-predicted micromotion and beyond-SMA macromotion appear simultaneously. These results point to a direction to investigate the quantum spin-mixing dynamics beyond the SMA, particularly in recent experiments of strongly interacting spin-1 atomic gases. 
		
	\end{abstract}
	
	\maketitle
	
	\section{Introduction}
	\label{sec:intro}
	Since their realization, optically trapped spinor gases with spin degrees of freedom have been among the most actively studied systems within the ultracold realm~\cite{Stamper-Kurn1998OpticalConfinementBoseEinstein, Ho1998SpinorBoseCondensates, Ohmi1998BoseEinsteinCondensationInternal, Kawaguchi2012SpinorBoseEinsteina, Stamper-Kurn2013SpinorBoseGases}. Combined with optical lattices and microwave dressing fields, spinor gases offer an unprecedented degree of control over many parameters, providing a powerful platform to investigate spin effects in conventional condensed matter physics~\cite{Zhao2014DynamicsSpinorCondensates, Zhao2015AntiferromagneticSpinorCondensates, Dag2018ClassificationQuenchdynamicalBehaviors}. Current research efforts are directed toward many hot topics, including novel quantum phase transitions~\cite{Gu2003FerromagneticPhaseTransition, Zhou2010SpinDynamicsDomain, Evrard2021ObservationFragmentationSpinor}, quantum magnetism~\cite{Vinit2013AntiferromagneticSpatialOrdering, Higbie2005DirectNondestructiveImaging}, spin-orbit coupling effects~\cite{Wang2010SpinOrbitCoupledSpinor, Lan2014RamandressedSpin1Spinorbitcoupled}, spin-mixing dynamics~\cite{Law1998QuantumSpinsMixing, Zhang2005CoherentSpinMixing, Widera2005CoherentCollisionalSpin}, and quantum metrology beyond the standard quantum limit~\cite{Cao2025JointEstimationTwoPhase,Liu2022NonlinearInterferometryClassical}, etc.

	The single-mode approximation (SMA) has benefited researchers in investigations of simplified quantum spin dynamics and coherent spin dynamics within the mean-field framework in ultracold spinor gases~\cite{Ho1998SpinorBoseCondensates,Ohmi1998BoseEinsteinCondensationInternal,Zhang2003MeanFieldGround}. However, recent experiments in a strongly interacting ferromagnetic $^{7}\text{Li}$ spinor condensate and an antiferromagnetically interacting $^{23}\text{Na}$ condensate exhibit novel multi-spatial-mode observations~\cite{Huh2020ObservationStronglyFerromagnetic, Jie2023DynamicalMeanfielddrivenSpinorcondensate}, triggering the exploration of mean-field and quantum spinor theory beyond the SMA~\cite{Jie2020MeanfieldSpinoscillationDynamics}.

	In this work, we consider two spin-1 atoms of $^{7}\text{Li}$, $^{87}\text{Rb}$, or $^{23}\text{Na}$ confined in a one-dimensional (1D) hard-wall box trap, focusing on the parameter regime beyond the SMA. Highly anisotropic potential traps have been realized in low-dimensional quantum gases, where many-body physics was extensively investigated~\cite{Paredes2004TonksGirardeauGas, Armijo2010ProbingThreeBodyCorrelations}. Recent advances in optical potential engineering have created even  uniform box traps~\cite{Jaksch1998ColdBosonicAtoms,VanEs2010BoxTrapsAtom, Tajik2019DesigningArbitraryOnedimensional, Chomaz2015EmergenceCoherenceTransverse, Gaunt2013BoseEinsteinCondensationAtoms}, which enable fine comparison between many-body quantum theories and experimental results. With this technique, homogeneous BECs of trapped bosonic atoms in quasi-one dimension~\cite{VanEs2010BoxTrapsAtom, Tajik2019DesigningArbitraryOnedimensional}, two dimensions~\cite{Chomaz2015EmergenceCoherenceTransverse}, and three dimensions~\cite{Gaunt2013BoseEinsteinCondensationAtoms} have been realized.
	
	For the many-body quantum system focused on in this work, a core theoretical challenge is how to obtain accurate physical properties when the Hilbert space dimension grows exponentially with the system size. To overcome this difficulty, we employ the generalized Van Vleck (GVV) perturbation theory~\cite{VanVleck1929TypeDoublingElectron, Jordahl1934EffectCrystallineElectric}, a canonical perturbative scheme that systematically reduces the problem to an effective Hamiltonian in a drastically smaller subspace, while faithfully preserving the low-energy spectrum. For two bosons in a 1D box, achieving a relative energy error of $10^{-6}$ through traditional exact diagonalization (ED) requires a Hamiltonian matrix with dimensions on the order of hundreds of millions. However, by adopting GVV, we find that the Hamiltonian matrix dimensions are reduced amazingly to a thousand-scale size while still maintaining the same relative error.

The present work aims to clarify the validity and breakdown of the SMA from a few-body perspective. Although SMA has been widely employed in the theory of spinor gases, its validity becomes questionable when the quadratic Zeeman energy becomes comparable to the spatial excitation energy and the spin-dependent interaction energy. ﻿In this regime, quadratic-Zeeman-induced spin redistribution couples spin dynamics to spatial excitations, giving rise to phenomena beyond the SMA.

	To explore the quantum spinor system beyond the SMA, we utilize the accurate GVV method to investigate ground-state properties and quench spin dynamics of $^{7}\text{Li}$, $^{87}\text{Rb}$, and $^{23}\text{Na}$ atoms. By emphasizing experiment-related observables, we calculate the differences among the density distributions of spin components for the ground state in a variety of magnetic fields, which directly reveal whether the SMA is broken. For quenching spin dynamics by suddenly changing the quadratic Zeeman energy $q$, we search for additional dynamical signals, such as additional resonances, beyond the SMA's prediction. In this manner, we are able to estimate the parameter regimes where the SMA becomes invalid. Our numerical results indicate that in a large-$q$ regime the SMA is violated in general. Obviously, these findings are ready to be verified in experiments of ultracold spin-1 atomic gases.

	\section{Model and Method}
	\label{sec:model}
	
	\subsection{Hamiltonian}
	\label{subsec:hamiltonian}
	The many-body Hamiltonian describing a 1D ultracold spin-1 Bose gas in a hard-wall box trap with $N$ atoms of mass $M$ is given by
	\begin{equation}
		\label{eq1}
		\begin{split}
			H =  &\sum_{i=1}^{N} \Bigl( -\frac{\hbar^{2}}{2M} \frac{\partial^{2}}{\partial x_{i}^{2}} - p f_{z,i} + q f_{z,i}^{2} \Bigr) \\
			& + \sum_{i<j} \delta(x_i - x_j) \bigl( c_0 + c_2 \vec{f}_i \cdot \vec{f}_j \bigr)
		\end{split}
	\end{equation}
	where $\vec{f}_i = (f_{x, i}, f_{y, i}, f_{z, i})^T$ is the spin-1 operator vector for the $i$th atom, and $c_0={2\hbar^2(a_0 + 2a_2)}/(3Ma_\perp^2)$ and $c_2={2\hbar^2(a_2- a_0)}/(3Ma_\perp^2)$ correspond, respectively, to the spin-independent and spin-dependent effective interaction coefficients~\cite{Mistakidis2023FewbodyBoseGases}. Here, $a_{0}$ and $a_{2}$ are the three-dimensional (3D) $s$-wave scattering lengths of the atoms in the scattering channels with total spin $F=0$ and $F=2$, respectively. The box length is $L$ ($0 \le x_i \le L$). The transversal confinement length scale is $a_{\perp}=\sqrt{\hbar /(M \omega_{\perp})}$ by assuming a large transversal trapping frequency $\omega_{\perp}$ such that $a_\perp \ll L$. The parameters $q$ and $p$ describe, respectively, the quadratic and linear Zeeman energy shifts. In this work, we consider the cases of $^{7}\text{Li}$, $^{87}\text{Rb}$ and $^{23}\text{Na}$ atoms with $N=2$ and zero total magnetization $\sum_i f_{z,i} = 0$. In this fixed-magnetization subspace, the linear Zeeman term gives only a constant energy shift, which can be neglected. For simplicity, we set $\hbar=1, M=1, L=1$, and $c_0=1$ in the following. The values of $c_2/c_0$ are approximated as $-1/2$, $-1/200$, and $3/100$ for $^{7}\text{Li}$, $^{87}\text{Rb}$, and $^{23}\text{Na}$, respectively~\cite{Stamper-Kurn2013SpinorBoseGases, Huh2020ObservationStronglyFerromagnetic}.

	\subsection{Generalized Van Vleck perturbation theory}
	\label{subsec:gvv}
	Using the Bethe ansatz (BA) hypothesis, Cao et al. solved a 1D model of interacting spin-1 bosons under the periodic boundary condition with antiferromagnetic spin-dependent interaction for $c_2 = c_0 > 0$~\cite{Cao2007PairedStateIntegrable}. However, exact solutions via the BA are not available other than this special integrable point. Instead of seeking an exact analytical solution, we therefore resort to ​​ED in the noninteracting boson basis. The numerical accuracy is controlled by the basis cutoff.

	\begin{figure}
		\centering
		\includegraphics[width=8.6cm]{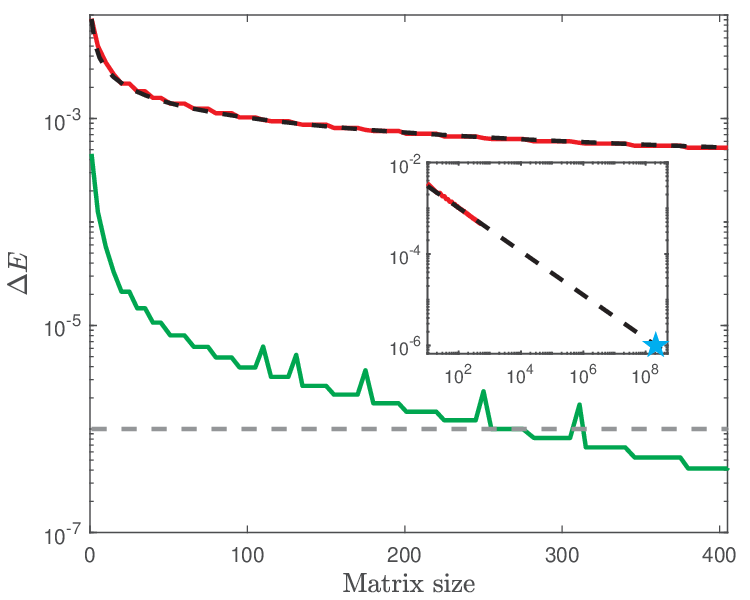}
		\caption{
			\label{FIG1} 
			Convergence of the relative error of the ground-state energy, $\Delta E$, for two atomic bosons in a one-dimensional hard-wall trap, computed as a function of the matrix dimension in the truncated basis. The red solid line shows the relative error obtained from ED, and the black dashed line represents the corresponding fit using the fitting function $a x^k$ ($a \approx 0.009$ and $k \approx -0.48$). The horizontal gray dashed line represents the target accuracy of $10^{-6}$. The green solid line shows the result of the GVV method, which achieves the target accuracy at a matrix dimension of $315$. The inset illustrates the extrapolation of the fitting curve. The cyan star marks where the relative error of ED reaches $10^{-6}$, which requires a Hilbert-space dimension of $\mathcal{O}(10^8)$. This benchmark demonstrates the computational efficiency of the GVV method and provides a quantitative convergence check in the exactly solvable spinless limit. 
		}
	\end{figure}

	To check the validity of the ED numerical method, we consider fully polarized spin-1 atoms, which are actually scalar or spinless atoms. Here, two Zeeman shifts $p$ and $q$ are both constant and neglected. The model was analytically solved by Gaudin in the early 1970s~\cite{Gaudin1971BoundaryEnergyBose}. For the conventional truncation-based ED, it suffers from slow convergence: achieving a relative error $\Delta E = |E-E_B|/|E_B|$ of $10^{-6}$ for the ground-state energy $E$ (compared to Gaudin's BA solution $E_B$) requires a truncated Hilbert-space dimension as large as $\mathcal{O}(10^8)$, as illustrated in Fig.~\ref{FIG1}.

	To make the problem computationally tractable, we employ the GVV transformation~\cite{VanVleck1929TypeDoublingElectron, Jordahl1934EffectCrystallineElectric}, a canonical perturbative method that effectively decouples the relevant low-energy subspace $P$ from the complementary excluded space $Q$ via a unitary transformation. The GVV method eliminates off-diagonal couplings order by order according to perturbation theory, enabling an accurate description of the effective Hamiltonian within a significantly reduced low-energy subspace. As shown in Fig.~\ref{FIG1}, the GVV method reduces the matrix dimension required for achieving numerical precision on the order of $10^{-6}$ from $\mathcal{O}(10^8)$ in the ED of the fully polarized case to merely $\mathcal{O}(10^3)$. This result confirms that the low-energy physics of the spinless problem is accurately reproduced by the GVV effective Hamiltonian once the model space is sufficiently enlarged.

	We note that the GVV method is a perturbation construction and therefore has a finite range of validity. Its accuracy is controlled by the coupling between the retained space $P$ and the excluded space $Q$, relative to the corresponding unperturbed energy difference~\cite{Bravyi2011SchriefferWolffTransformation, Devarajan2008GeneralizedVanVleck}. For the ground-state properties considered here, a useful estimate of the relevant perturbative parameter is
		\begin{equation*}
			\eta_g =
			\max_{\beta\in Q}
			\frac{|V_{g\beta}|}
			{|E^{(0)}_g-E^{(0)}_\beta|},
		\end{equation*}
	where $|g\rangle$ denotes the ground state in the retained $P$ space, $V_{g\beta}$ is its coupling to a state $|\beta\rangle$ in $Q$, and $|E^{(0)}_g-E^{(0)}_\beta|$ is the corresponding unperturbed energy difference. For the benchmark shown in Fig.~\ref{FIG1}, when the \(P\)-block size $N_P$ reaches $315$, we find $\eta_g\simeq 3.8\times10^{-4}$. This indicates that the eliminated $Q$ subspace is only weakly mixed with the target low-energy state. In the spin-1 calculations below, we therefore use a slightly larger spatial $P$ block with $N_P=400$. After taking into account the spin degree of freedom, bosonic permutation symmetry, and the zero-magnetization constraint, the effective Hamiltonian has a reduced matrix dimension of $1178$. For the spin-1 system considered here, the spin-dependent interaction \(c_2\) is much weaker than the spin-independent interaction \(c_0\), and it does not qualitatively change the separation between the low-energy spatial sector and the excluded high-energy sector. This provides a controlled basis for applying the GVV method to the low-energy properties considered in this work. The GVV treatment may become less accurate when the $P$-$Q$ coupling becomes comparable to the energy difference, for example the violation of the condition $\eta_g \ll 1$ if one chooses a rather small $P$-block size.

	\begin{figure*}[t!] 
		\centering
		\includegraphics[width=17.2cm]{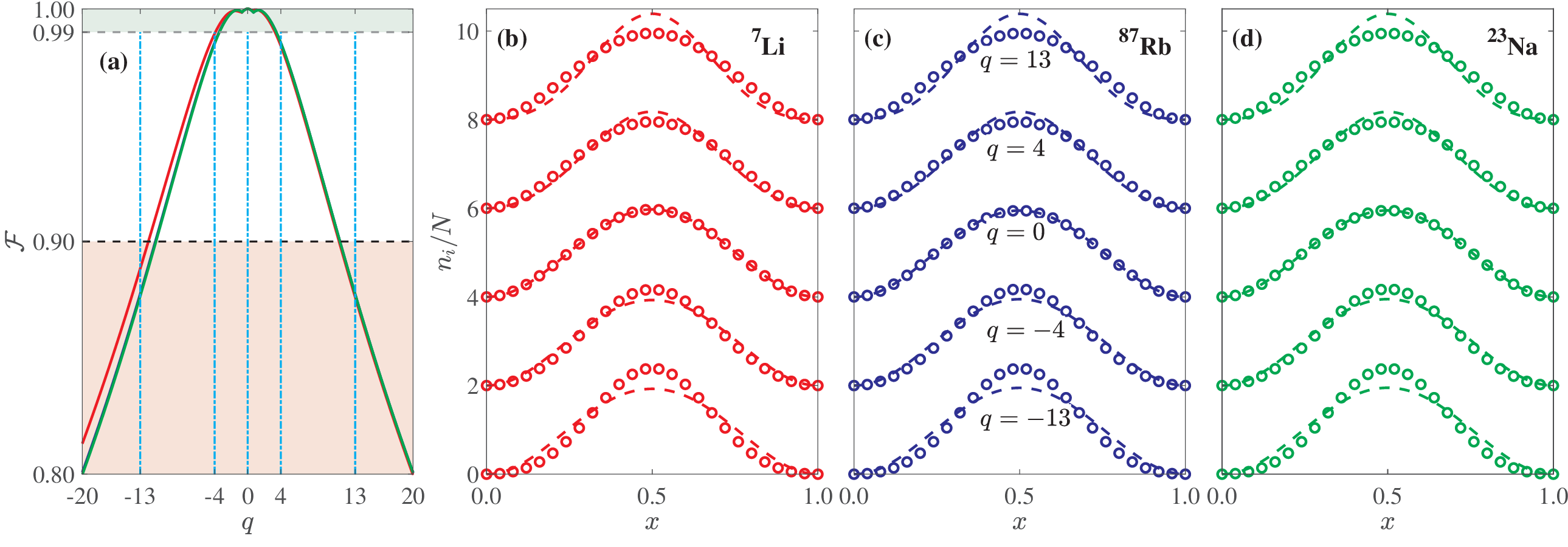}
		\caption{
			\label{FIG2}
			(a) Fidelity $\mathcal{F}$ between spin component $m_F=0$ and $m_F=1$ as a function of $q$, for three different spin-dependent interaction strengths $c_2$, labeled as $^{7}\text{Li}$ (red), $^{87}\text{Rb}$ (blue), and $^{23}\text{Na}$ (green). The three curves overlap strongly and are hardly distinguishable for $q>0$. The shaded regions indicate three regimes defined
			by the fidelity: the SMA-valid regime with \(\mathcal F\ge0.99\), the crossover regime with \(0.9\le\mathcal F<0.99\), and the beyond-SMA regime with \(\mathcal F<0.9\). Two horizontal gray dashed lines mark $\mathcal F = 0.99$ and $\mathcal F = 0.9$. The vertical cyan dashed lines are guides at $q = 0, \pm 4, \pm 13$. Spatial density distribution $n_{m_F}$ of spin component $m_F = 0$ (circles) and $m_F = 1$ (dashed lines) for $^{7}\text{Li}$ (b), $^{87}\text{Rb}$ (c), and $^{23}\text{Na}$ (d) at $q = -13, -4, 0, 4, 13$ (from bottom to top). For clarity, the lines are shifted upward by an increment of $2$. One can immediately find that the SMA is broken at large $|q|$.
		}
		\vspace*{-3mm} 
	\end{figure*}
	
	\section{Ground-State Properties}
	\label{sec:ground}
	With the efficient and accurate GVV numerical method, we calculate the ground-state properties of two spin-1 atoms for various $q$, focusing on the validity of the SMA. To quantify the validity of the SMA, we introduce quantum fidelity defined as
	$\mathcal{F} (\rho_0, \rho_1) = \left[ \operatorname{Tr} \left( \sqrt{\sqrt{\rho_0} \rho_1  \sqrt{\rho_0  }} \right) \right]^2$~\cite{Jozsa1994FidelityMixedQuantum,Nielsen2010QuantumComputationQuantum}, where $\rho_{m_F}(x, x')$ ($m_F=0, \pm 1$) are normalized single-particle reduced density matrices ($\operatorname{Tr} \left(\rho_{m_F}\right) = 1$). Such a fidelity ranges from 0 to 1. The higher the fidelity, the more valid the SMA. For the convenience of experimentalists, we also calculate the normalized spatial density distribution of each spin component, $n_{m_F} (x)=\rho_{m_F} (x,x)$ ($m_F = 0, \pm 1$), which is easier to observe than the fidelity in experiments.

	In Fig.~\ref{FIG2} we summarize the ground-state properties as a function of the quadratic Zeeman shift $q$, where $N = 2$. Figure~\ref{FIG2}(a) displays the fidelity $\mathcal{F}$ for $^{7}\text{Li}$, $^{87}\text{Rb}$, and $^{23}\text{Na}$. We classify the parameter space into three regimes: the SMA-valid regime ($\mathcal F\ge0.99$), the crossover regime ($0.9\le\mathcal F<0.99$), and the beyond-SMA regime ($\mathcal F<0.9$). In all cases, $\mathcal{F}$ decreases apparently with $|q|$, indicating a gradual reduction in spatial overlap between the spin components in a strong magnetic field, i.e., a gradual breakdown of the SMA. Such a clear dependence suggests that we ascribe the breakdown of the SMA to the competition between the quadratic Zeeman energy and the kinetic energy. We note that this result for two atoms is different from previous understanding for many atoms, which is often stated as the violation of the condition $|c_2| \ll c_0$~\cite{Law1998QuantumSpinsMixing, Pu1999SpinmixingDynamicsSpinor, Yi2001TrappedCondensatesAtoms, Zhang2003MeanFieldGround}.

	In fact, the fidelity exhibits a very weak dependence on $c_2$ and is insensitive to its sign, as shown in Fig.~\ref{FIG2}(a). The analysis of the perturbation theory confirms the numerical results and further identifies the origin of the observed $c_2$ dependence (see Appendix~\ref{sec:perturbation}). Specifically, for $q>0$, the $\rho_{m_F}$ are essentially independent of $c_2$, resulting in almost complete overlap of the fidelity curves for different $c_2$ values. For $q<0$, the more populated spin component depends on $c_0-c_2$, leading to a separation of the fidelity curves, with larger $|c_2|$ corresponding to a slightly faster decay of the fidelity. This weak $c_2$ dependence justifies using Fig.~\ref{FIG2}(a) as an effective one-dimensional regime diagram for the parameter range considered here.

	To reveal the close connection between the fidelity and the spatial structure of spin components, we use the two boundaries $F=0.99$ and $F=0.9$, marked by horizontal dashed lines in Fig.~\ref{FIG2}(a), and select representative values $q = \pm 4$ and $q = \pm 13$ around these intersections, together with $q=0$. We plot in Figs.~\ref{FIG2}(b)-\ref{FIG2}(d) the single-particle density distribution $n_{m_F} (x)$ ($m_F = 0,1$) for $^{7}\text{Li}$, $^{87}\text{Rb}$, and $^{23}\text{Na}$, respectively. At $q = 0$, all three systems exhibit nearly identical spatial distributions, consistent with the high fidelity observed in Fig.~\ref{FIG2}(a). As $|q|$ increases to $q = \pm 4$, small deviations emerge between the two spin components. At $q = \pm 13$, density distributions show clear distinction for the two spin components in all three systems, confirming the breakdown of the SMA.

	Although the SMA is broken at a large enough $|q|$, as shown in Fig.~\ref{FIG2}, the spatial structure of the density distributions is rather different from previous beyond-SMA mean-field results. In mean-field theory, the more populated spin component ($m_F = 0$ for a large and positive $q$ and $m_F = +1$ for a large and negative $q$) occupies the center region and the less populated component is pushed away from the center, exhibiting a phase separation~\cite{Isoshima1999SpindomainFormationSpinor, Stenger1998SpinDomainsGroundstate, Matuszewski2009ExcitedSpinStates, JimenezGarcia2019SpontaneousFormationRelaxation, Chang2005CoherentSpinorDynamics,Zhang2005DynamicalInstabilityDomain, Jie2020MeanfieldSpinoscillationDynamics, Jie2023DynamicalMeanfielddrivenSpinorcondensate,Huh2020ObservationStronglyFerromagnetic,Kim2021EmissionSpinCorrelatedMatterWavea}. In this work, however, the density is high near the center and low near the boundaries for all components, implying the mixing nature of spin components. This spatial structure is again consistent with the perturbative analysis (see Appendix~\ref{sec:perturbation}), which shows that increasing $|q|$ induces coherent contributions from high-energy spatial modes, gives rise to constructive interference at the trap center, and results in the observed enhancement of the less populated spin-component density.

	\section{Quench Spin Dynamics }
	\label{dynamics}
	
	In this section, we investigate the dynamical evolution process, restricting our analysis to positive values of $q$. The initial state is chosen as the ferromagnetic ground state at $q=0$, $|\Psi\rangle =\left(\sqrt{2}\,|\Phi_1\rangle + |\Phi_2\rangle\right)/\sqrt{3}$, where $|\Phi_1\rangle= |\varphi_1(x_1)\varphi_1(x_2)\rangle \otimes |0,0\rangle$ and $|\Phi_2\rangle =|\varphi_1(x_1)\varphi_1(x_2)\rangle \otimes \left(|1,-1\rangle + |-1,1\rangle\right)/\sqrt{2}$, with $|\varphi_n(x)\rangle$ denoting the single-particle eigenstates of the 1D hard-wall box trap. For this initial state, $p_0 = 2/3$ and $p_{\pm 1} = 1/6$, with $p_{m_F}$ denoting the population fraction of the spin component $m_F$. The system starts to evolve after $q$ is suddenly quenched to a positive value. In experiments, the dynamical quench of $q$ can be realized by abruptly changing the bias magnetic field, since the magnetic contribution to the quadratic Zeeman shift scales as $q_B/h\propto B^2$. Alternatively, an off-resonant microwave dressing field can be used to generate an additional tunable quadratic Zeeman shift~\cite{Zhao2014DynamicsSpinorCondensates,Li2020ManipulationHeteronuclearSpin}.
	
	\begin{figure*}[t!] 
		\centering
		\includegraphics[width=17.2cm]{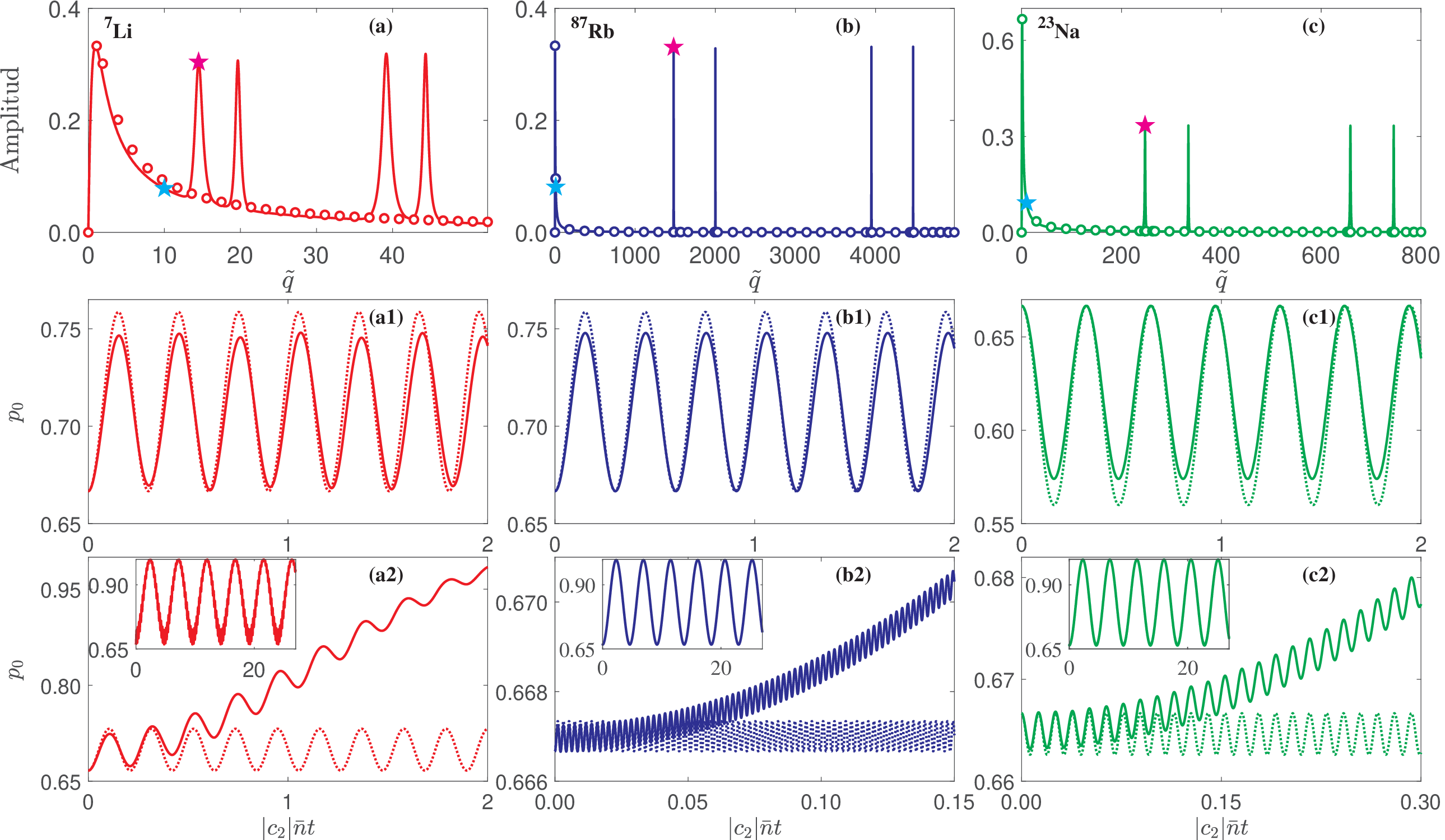}
		\caption{
			\label{FIG3}
			Spin-mixing dynamics predicted by the GVV method and the SMA. Top row: oscillation amplitude of $p_{0}(t)$ as a function of $\tilde{q}$ for $^{7}\text{Li}$ (a), $^{87}\text{Rb}$ (b), and $^{23}\text{Na}$ (c). Multiple peaks appear with the GVV method (solid lines) but are absent under the SMA (circles). Middle row: evolution of $p_{0}(t)$ at $\tilde q = 10$ (cyan stars in (a--c)) for $^{7}\text{Li}$ (a1), $^{87}\text{Rb}$ (b1), and $^{23}\text{Na}$ (c1). At this value of $\tilde q$, the evolution predicted by the GVV method (solid lines) agrees well with that predicted by the SMA (dashed lines). Bottom row: $p_0(t)$ at $\tilde{q}=14.5$ for $^{7}\text{Li}$ (a2), $1481.2$ for $^{87}\text{Rb}$ (b2), and $247.2$ for $^{23}\text{Na}$ (c2), corresponding to magenta stars in (a--c), respectively. At these values of $\tilde q$, the GVV method (solid lines) predicts short-period micromotion and long-period macromotion, while the SMA (dashed lines) captures only the micromotion. The insets in panels (a2--c2) illustrate the long-time macromotion with the GVV method.
		}
		\vspace*{-3mm} 
	\end{figure*}

	We present in Fig.~\ref{FIG3} a systematic comparison of $^{7}\text{Li}$, $^{87}\text{Rb}$, and $^{23}\text{Na}$ systems obtained numerically by the GVV method and the SMA. We choose $\varphi_1(x)$ as the spatial mode under the SMA. The top row quantifies the oscillation amplitude of $p_{0}(t)$ as a function of the dimensionless quadratic Zeeman shift $\tilde{q} \equiv q/(|c_2| \bar n)$ for $^{7}\text{Li}$, $^{87}\text{Rb}$, and $^{23}\text{Na}$, where $\bar n \equiv N/L=2$. The SMA predicts a single peak in amplitude with increasing $\tilde{q}$, while the GVV results exhibit additional multiple peaks. The first peak at low $\tilde q$ with a long tail is qualitatively captured both by the SMA and the GVV method. Their values agree well with each other. The additional peaks at large $\tilde q$, exhibiting almost the same peak height for each case, are predicted by the GVV method but missed by the SMA. Apparently, these additional peaks imply the failure of the SMA in the region of large $\tilde q$.

	To highlight the detail of the spin dynamics, two representative values of $\tilde q$, marked by stars, are chosen. One is $\tilde q = 10$ where the GVV and SMA results agree quantitatively, and the other is at the second peak where the GVV amplitude is strongly enhanced, namely, $\tilde{q}=14.5$ for $^{7}\text{Li}$, $1481.2$ for $^{87}\text{Rb}$, and $247.2$ for $^{23}\text{Na}$. The middle row in Fig.~\ref{FIG3} shows the time evolution of $p_{0}(t)$ at $\tilde{q}=10$ for all three cases. The SMA results remarkably agree with the GVV for $^{87}\text{Rb}$, $^{7}\text{Li}$, and $^{23}\text{Na}$. In all three cases, the GVV and SMA results share nearly identical single-frequency oscillations with only slight quantitative differences, i.e., the GVV dynamics consistently yields a little smaller oscillation amplitude compared to the SMA. Such agreements indicate that the SMA provides a quantitatively reliable description of the spin dynamics in this low-$\tilde q$ regime.
	
	As a sharp contrast, in the bottom row of Fig.~\ref{FIG3} (corresponding to the second resonant point), the GVV results display a qualitatively different behavior from the SMA. At this resonance, the SMA continues to predict a simple single-frequency oscillation. However, the oscillation from the GVV method exhibits a complex beat pattern. The short period and small amplitude oscillation is similar to the micromotion in trapped ions, while the long period and large-amplitude oscillation is similar to the macromotion~\cite{Leibfried2003QuantumDynamicsSinglea,Paul1990ElectromagneticTrapsChargedb}. The coexistence of the micro- and macromotion indicates the emergence of a double-frequency structure. To illustrate this macromotion more clearly, the GVV results are displayed over a much longer timescale in the insets. Although the SMA predicts successfully the short-period, small-amplitude micromotion, it fails to catch the long-period, large-amplitude macromotion.

	\begin{table}[t] 
		\centering 
		\small 
		\setlength{\tabcolsep}{5pt} 
		\renewcommand{\arraystretch}{1.15} 
		\caption{
			Dominant ordinary frequencies extracted from the GVV and SMA dynamics at the second resonance in Fig.~\ref{FIG3}. 
			The Fourier transform is performed with respect to the scaled time \(\tau=t|c_2|\bar n\) in the numerical units with \(\hbar=1\). 
			The listed frequencies are therefore dimensionless ordinary frequencies in the scaled time \(\tau\). 
			The low-frequency component \(\nu_M\) corresponds to the macromotion. 
			For the GVV high-frequency doublet, \(\nu_{\mu,c}^{\rm GVV}\) denotes the center frequency and \(\delta\nu_\mu\) denotes the half splitting. 
			The last column gives the large-\(q\) estimate for the SMA-like micromotion carrier frequency.
		} 
		\label{tab:freq} 
		\begin{tabular}{@{}lccccc@{}} 
			\toprule  
			& \(\tilde q_r\) 
			& \(\nu_{\mu,c}^{\rm GVV}\pm\delta\nu_\mu\) 
			& \(\nu_M^{\rm GVV}\) 
			& \(\nu^{\rm SMA}\) 
			& \(\tilde q_r/\pi\) \\ 
			\midrule 
			\(^{7}\mathrm{Li}\)  & 14.5   & \(4.72\pm0.10\)    & 0.20 & 4.74   & 4.62 \\ 
			\(^{87}\mathrm{Rb}\) & 1481.2 & \(471.58\pm0.11\)  & 0.22 & 471.59 & 471.48 \\ 
			\(^{23}\mathrm{Na}\) & 247.2  & \(78.59\pm0.11\)   & 0.22 & 78.56  & 78.69 \\ 
			\bottomrule 
		\end{tabular} 
	\end{table}

	To quantify the characteristic timescales observed at the second resonance, we perform a Fourier analysis of \(p_0(t)\) with respect to the scaled time \(\tau=t|c_2|\bar n\). The extracted dominant frequencies are summarized in Table~\ref{tab:freq}. In the GVV spectra, the high-frequency micromotion appears as a doublet. We therefore list its center frequency \(\nu_{\mu,c}^{\rm GVV}\) together with the half splitting \(\delta\nu_\mu\). The center frequency is close to the SMA frequency and to the large-\(q\) estimate \(\tilde q_r/\pi\), whereas the low-frequency component \(\nu_M^{\rm GVV}\), absent in the SMA, characterizes the macromotion. The relation \(2\delta\nu_\mu\simeq\nu_M^{\rm GVV}\) indicates that the high-frequency doublet splitting and the macromotion reflect the same resonant spin-spatial coupling beyond the SMA, as further explained by the three-level model below.

	To understand comprehensively the mechanism behind the complex spin micro- and macromotion, we propose an effective three-level model by truncating the full many-body Hamiltonian Eq.~(\ref{eq1}) to the subspace spanned by $\{|1\rangle, |2\rangle, |3\rangle\}$, see Appendix~\ref{sec:threelevel}. These states are given by $|1\rangle = |\varphi_1(x_1)\varphi_1(x_2)\rangle \otimes |0,0\rangle$, $ |2\rangle= |\varphi_1(x_1)\varphi_1(x_2)\rangle \otimes \left(|1,-1\rangle + |-1,1\rangle\right)/\sqrt{2}$, and $|3\rangle=\left(|\varphi_{n_i}(x_1)\varphi_{n_j}(x_2)\rangle+|\varphi_{n_j}(x_1)\varphi_{n_i}(x_2)\rangle \right)/\sqrt{2(1+\delta_{n_i,n_j})} \otimes |0,0\rangle $. For instance, at the second resonance, $n_i = n_j = 2$. Despite its simplicity, the three-level model captures the essential interplay between spin and spatial degrees of freedom and provides reasonable explanations for the key dynamical features observed with the GVV method.

	According to this model, the fast micromotion is associated with the states $|1\rangle$ and $|2\rangle$, which share the same spatial ground state. If the coupling to the spatially excited state \( |3\rangle \) is neglected, the corresponding angular frequency is $\omega_\mu\simeq\sqrt{(E_2-E_1)^2 + 4V_{12}^2}$, where the energy difference $|E_2-E_1|$ is controlled by $2q$, and the spin dependent interaction $V_{12}$ is proportional to $|c_2|\bar n$. In the large-$q$ regime, $E_2-E_1\simeq 2q$, and therefore the ordinary frequency measured with respect to the scaled time $\tau=t|c_2|\bar n$ is
		\[
		\nu_\mu \simeq \frac{1}{2\pi}\frac{2q}{|c_2|\bar n}
		= \frac{\tilde q}{\pi}.
		\]
		This explains why the SMA frequency \(\nu^{\rm SMA}\) is close to the large-\(q\) estimate \(\tilde q_r/\pi\) in Table~\ref{tab:freq}.
	
	Near a resonant point \(\tilde q_r\), however, \( |2\rangle \) becomes nearly degenerate with the spatially excited state \( |3\rangle \). The resonant coupling \(V_{23}\) hybridizes \( |2\rangle \) and \( |3\rangle \), opens an avoided-crossing gap, and forms two dressed states \( |+\rangle \) and \( |-\rangle \). As a result, the original SMA-like high-frequency component associated with \( |1\rangle\leftrightarrow |2\rangle \) becomes a doublet in the GVV spectrum. The center of this doublet remains close to the SMA frequency, while its splitting is governed by the same avoided-crossing gap that produces the low-frequency macromotion. This gives the approximate relation $2\delta\nu_\mu \simeq \nu_M$, consistent with Table~\ref{tab:freq}. The carrier frequency and the decreasing amplitude of the fast micromotion are therefore SMA-like. 
	
	The same avoided crossing also gives rise to the slow macromotion. Its characteristic angular frequency can be estimated as $\omega_M\simeq\sqrt{(E_3-E_2)^2+4|V_{23}|^2}.$ At exact resonance, \(E_2\simeq E_3\), this reduces to \(\omega_M\simeq 2|V_{23}|\). The corresponding ordinary frequency measured with respect to the scaled time is approximately
		\[
		\nu_M\simeq \frac{1}{2\pi}\frac{2|V_{23}|}{|c_2|\bar n}.
		\]	
	since \(V_{23}\) is proportional to \(|c_2|\bar n\), with a proportionality coefficient determined by the spin matrix element and the spatial wave-function overlap, \(\nu_M\) is expected to be nearly independent of atomic species in the scaled time, consistent with Table~\ref{tab:freq}. The associated oscillation amplitude exhibits a significant enhancement, reflecting the collective involvement of both spin and spatial excitation beyond the SMA. This resonant macromotion is similar to earlier research on beyond-SMA spin dynamics~\cite{Jie2020MeanfieldSpinoscillationDynamics, Jie2023DynamicalMeanfielddrivenSpinorcondensate}, which is in the mean-field theory. Additionally, although tuning $\tilde q$ can bring the state $|2\rangle$ into degeneracy with different spatially excited states, only some of them lead to observable resonant dynamics. This is because the spin interaction $V_{23}$ has selection rules from the overlap of spatial wave functions. As a result, the enhancement of the oscillation amplitude of $p_0(t)$ occurs only at certain specific and discrete values of $\tilde q$, giving rise to multiple resonant peaks.

	To connect with experimentally relevant parameters, we further estimate the resonant quadratic Zeeman energy using length scales realized in one-dimensional atom-chip Bose-gas experiments. Taking $L=160~\mu{\rm m}$ reported in Ref.~\cite{Tajik2019DesigningArbitraryOnedimensional}, the second-resonance condition gives 
		\begin{equation} 
			\frac{q_r}{h} \simeq \frac{3\pi^2}{2} \frac{\hbar^2}{M L^2 h}, \label{eq:qr_physical} 
		\end{equation} 
	for this length scale, one obtains
		\begin{equation*} 
			\frac{q_r}{h}\simeq
			0.835~{\rm Hz},\ 0.067~{\rm Hz},\ 0.254~{\rm Hz}
		\end{equation*} 
	for $^{7}\mathrm{Li}$, $^{87}\mathrm{Rb}$, and $^{23}\mathrm{Na}$, respectively. Equivalently, using $k_B T/h\simeq20.84\,{\rm Hz}\times T/(1\,{\rm nK})$, these values correspond to $q_r/k_B\simeq 40.1\,{\rm pK},\ 3.2\,{\rm pK},\ 12.2\,{\rm pK}$. Within this estimate, the corresponding spatial excitation gap is approximately $\Delta_{\mathrm{sp}}/h \simeq 2q_r/h$. As a representative spinor-BEC temperature, we take $T\simeq50~{\rm nK}$~\cite{Chang2005CoherentSpinorDynamics}, corresponding to $k_B T/h\simeq1.04~{\rm kHz}$. Thus, for $L=160~\mu{\rm m}$, the thermal energy scale is much larger than the axial spatial excitation gap associated with the second resonance.
		
	Nevertheless, finite temperature does not necessarily destroy spin dynamics. Coherent spin-population oscillations have been observed in multi-spatial-mode thermal spinor gases, including spin-1 $^{23}\mathrm{Na}$ and spin-1/spin-2 $^{87}\mathrm{Rb}$ gases~\cite{Pechkis2013SpinorDynamicsAntiferromagnetic, He2015CoherentSpinmixingDynamics}, where the thermal energy is much larger than the spin-dependent energy. The resonance predicted here is more demanding because it involves an avoided crossing between spin and spatial excitations beyond the SMA, similar to recent experimental observations that spatial mean-field dynamics can strongly affect spinor-condensate oscillations beyond the SMA~\cite{Jie2023DynamicalMeanfielddrivenSpinorcondensate}.  Therefore, these results indicate that a thermal energy scale larger than the spin-dependent interaction scale does not by itself preclude the spin dynamics.

	\section{Conclusions}
	In summary, we apply the accurate GVV perturbation theory to numerically explore the ground-state and quench spin dynamics of two spin-1 bosons in a 1D box trap by tuning the quadratic Zeeman shift $q$, focusing on the beyond-SMA regime. The ground-state fidelity decreases gradually as $|q|$ increases and the density distribution of the less populated spin component(s) deviates noticeably from that under the SMA at large $|q|$. After quenching $q$ from zero to a certain resonance value, the spin dynamics exhibits the SMA-predicted micromotion (rapid and small oscillation), as well as the beyond-SMA macromotion (slow and large oscillation). In addition, the ground-state and the quench spin dynamics can be well understood qualitatively by the standard perturbation theory and the three-level model, respectively. Given recent advances in ultracold spinor condensates, especially in $^7$Li atomic gases, the quantum spin mixing beyond the SMA by tuning the quadratic Zeeman shift is worth exploring experimentally in the near future~\cite{Huh2020ObservationStronglyFerromagnetic, Kim2021EmissionSpinCorrelatedMatterWavea}.

	\begin{figure*}[t!] 
		\centering
		\includegraphics[width=17.2cm]{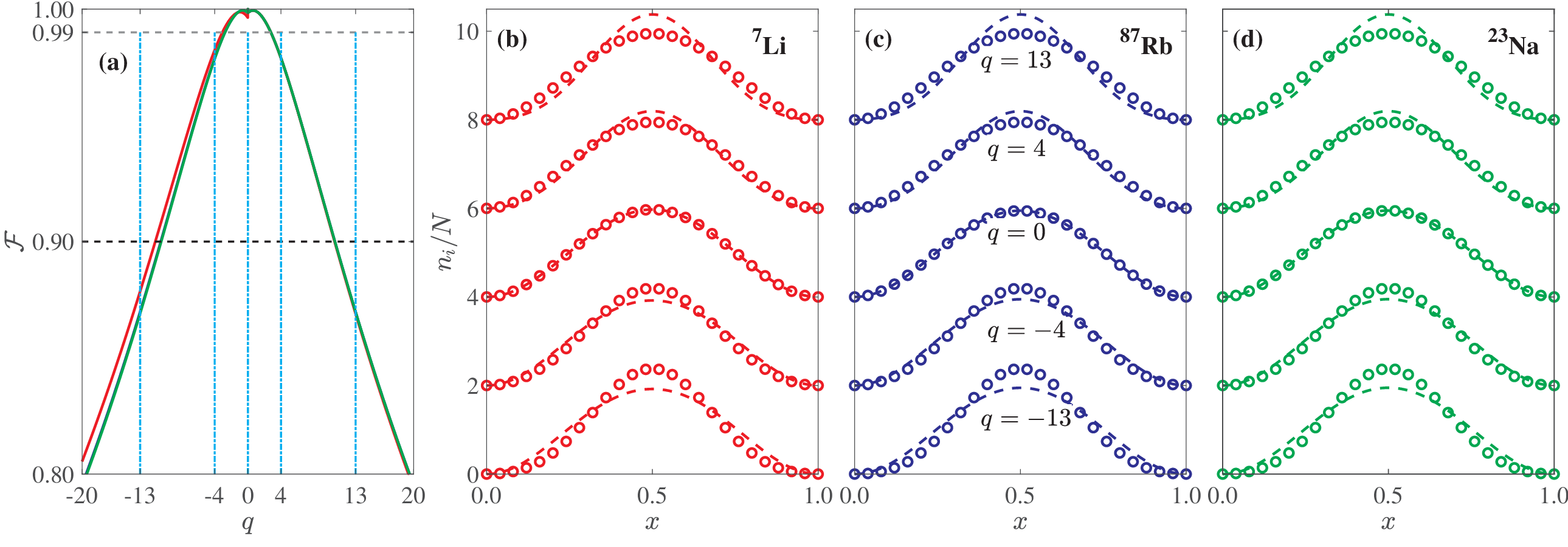}
		\caption{		
			Similar to Fig.~\ref{FIG2} in the main text, except that the data are calculated by the perturbation theory. 
		}
		\label{FIG4}
		\vspace*{-3mm} 
	\end{figure*}
	
	We now discuss the generalizability of the present method. GVV methods have been applied in quantum-chemistry treatments of correlated many-electron systems~\cite{Kirtman1981SimultaneousCalculationSeveral,Redmon1982MultidimensionalManybodyTheory}. Thus, in principle, the present framework can be generalized to more complex many-body systems once a suitable low-energy model space $P$ is identified. The main practical challenge is the rapid growth of the many-body Hilbert space with particle number and with the number of retained spatial modes. 
		
	The validity of the SMA is expected to be controlled by the competition between the spatial excitation gap and the spin-dependent energy (including the spin exchange interaction with $c_2$ and the quadratic Zeeman shift). Increasing the particle number can make the SMA less reliable by enhancing the density-dependent spin exchange interaction scale or by reducing the lowest spatial excitation gap in larger systems. This trend is consistent with previous studies~\cite{Yi2002SinglemodeApproximationSpinor1, Jie2020MeanfieldSpinoscillationDynamics, Jie2023DynamicalMeanfielddrivenSpinorcondensate}. 
		
	The present work is based on the conventional two-body spin-mixing Hamiltonian for dilute spinor gases~\cite{Stamper-Kurn2013SpinorBoseGases}. Interactions beyond two-body collisions may also influence the validity of the SMA. While weak or spin-independent three-body collisions are expected to mainly renormalize the effective interaction parameters, coherent spin-dependent or effective three-body interactions can generate additional couplings between spin and spatial degrees of freedom, thereby reducing the validity of the SMA when their characteristic energy scale becomes comparable to the relevant spatial excitation gap (particularly at particle densities above $10^{17}$ cm$^{-3}$)~\cite{Mestrom2021ThreebodySpinMixing}.

\begin{acknowledgments}
		W.Z. thanks Xiwen Guan, Jianwen Jie, and Qiang Gu for helpful discussions. The work is supported by the National Natural Science Foundation of China (NSFC) under Grants No. 12274331 and No. 92576103, Quantum Science and Technology - National Science and Technology Major Project (Grant No. 2025ZD0300800), and the HZNU scientific research and innovation team project (Grant No. TD2025003). The numerical calculations in the paper have been partially done on the supercomputing system in the Supercomputing Center of Wuhan University.
\end{acknowledgments}

	\begin{appendix}
		
		\section{Perturbative analysis }
		\label{sec:perturbation}

		We provide an independent perturbative analysis that offers a  physical interpretation for the observations in ground-state properties. We rewrite​ the Hamiltonian for a 1D spin-1 Bose gas confined in a hard-wall box trap with $N$ atoms of mass $M$ as
		\begin{equation}
			\begin{split}
				H = H_0 + H'
			\end{split}
			\label{H_Definition}
		\end{equation}
		with
		\begin{align}
			&H_0 = \sum_{i=1}^N\left[-\frac{\hbar^2}{2M}\frac{\partial^2}{\partial x_i^2} - p\, f_{z,i} + q\, f_{z,i}^2\right], \nonumber \\
			&H' = \sum_{i<j}\delta(x_i-x_j)\bigl(c_0 + c_2\,\mathbf f_i\!\cdot\!\mathbf f_j\bigr)\nonumber
		\end{align}
		where \( H_0 \) represents the non-interacting part and \( H' \) contains the interaction terms. The single-particle eigenfunctions and energies in the trap are
		\begin{align*}
			\varphi_n(x)=\sqrt{2}\sin(n\pi x),\qquad 
			\varepsilon_n=\frac{\pi^2}{2} n^2, \quad n=1,2,\dots
		\end{align*}
		It is easy to solve for two atoms in the trap if one ignores the interaction term $H'$. We treat $H'$ as a perturbation, which mixes the excited spatial modes into the ground state. We employ standard perturbation theory to calculate corrections to the ground state of $H_0$. From this correction, the unnormalized single-particle reduced density matrix (SPRDM) for a given spin component $m_F$ (with $m_F = 0, 1$) can be calculated and written as 
		
		\begin{widetext}
			\begin{equation}
				\begin{split}
					\rho_{m_{F}}^{(1)}(x,x') \approx &\; 2R_1^2\,\varphi_1(x)\varphi_1^*(x')  +\sqrt{2}\,R_1S_1\Bigl(\varphi_1(x)\varphi_3^*(x')+\varphi_3(x)\varphi_1^*(x')\Bigr) \\[2mm] 
					&  +2\sum_{n\ge2}R_n^2\,\varphi_n(x)\varphi_n^*(x') \\[2mm]
					&+\sqrt{2}\sum_{n\ge2}R_nS_n\Bigl(\varphi_n(x)\varphi_{n+2}^*(x') +\varphi_{n+2}(x)\varphi_n^*(x')\Bigr) \\[2mm]  
					&+\sqrt{2}\sum_{n\ge1}S_nR_{n+2}\Bigl(\varphi_n(x)\varphi_{n+2}^*(x') +\varphi_{n+2}(x)\varphi_n^*(x')\Bigr) \\[2mm]		
					&+\sum_{n\ge1}S_n^2\Bigl(\varphi_n(x)\varphi_n^*(x') +\varphi_{n+2}(x)\varphi_{n+2}^*(x')\Bigr) \\[2mm] 
					&+\sum_{n\ge1}S_nS_{n+2}\,\Bigl(\varphi_n(x)\varphi_{n+4}^*(x') +\varphi_{n+4}(x)\varphi_n^*(x')\Bigr)
				\end{split}
				\label{eqrho_general}
			\end{equation}
		\end{widetext}
		where the coefficients $R_n$ and $S_n$ are determined by the interaction parameters and $q$. For the more populated spin component, $m_F=0$ if $q > 0$ or $m_F=1$ if $q < 0$, these coefficients are independent of $q$ and are given by
		\begin{align}
			&R_1=1, \qquad \nonumber\\
			&R_n= -\frac{B}{2\varepsilon_1(n^2-1)},\qquad (n\ge2), \nonumber\\
			&S_n=\frac{B}{2\sqrt{2}\,\varepsilon_1 (n+1)^2}, \qquad (n\ge1) \nonumber
		\end{align}
		where $B=c_0$ if $m_F = 0$ or $B=c_0-c_2$ if $m_F = 1$. Consequently, the SPRDM of the more populated spin component is immune to $q$ at this order of perturbation theory. In contrast, for the less populated spin component, $m_F=1$ if $q > 0$ or $m_F=0$ if $q < 0$, we find
		\begin{align}
			&R_1=-\frac{3c_2}{2\sqrt{2} |q|}, \qquad \nonumber \\
			&R_n=-\frac{c_2}{\sqrt{2}\bigl[\varepsilon_1(n^2-1)+|q|\bigr]},\;(n\ge 2),\nonumber\\
			&S_n=\frac{c_2}{2\bigl[\varepsilon_1(n+1)^2+|q|\bigr]},\;(n\ge 1)\nonumber
		\end{align}
		which depend closely on $q$. Such a dependence explains why the single-particle spatial density $n_{m_F} (x)=  \rho_{m_{F}}(x,x)$ of the less populated spin component evolves significantly with increasing $|q|$. 
		
		After normalizing the above analytical expressions $\rho_{m_{F}}(x,x') = \rho_{m_F}^{(1)} (x,x')/\operatorname{tr} \left(\rho_{m_F}^{(1)} (x,x')\right)$, we compute the fidelity $\mathcal{F} (\rho_0, \rho_1) = \left[ \operatorname{tr} \left( \sqrt{\sqrt{\rho_0} \rho_1  \sqrt{\rho_0  }} \right) \right]^2$ and the spatial density distribution $n_{m_F} (x)$. The results are presented in Fig.~\ref{FIG4}, which are similar to Fig.~2 of the main text. 
		
		Figure~\ref{FIG4}(a) demonstrates that the perturbative calculation reproduces all key features in the main text, (i) the decay of $\mathcal{F}$ with $|q|$ , (ii) the significant overlap of curves for different $c_2$ values for $q>0$, and (iii) the intersection points with the reference lines near $q \approx \pm 4$ and $q \approx \pm13$. Again, density distributions displayed in Fig.~\ref{FIG4}(b-d) match precisely the main text findings, (i) identical distributions for $n_0(x)$ and $n_1(x)$ at $q=0$ ($\mathcal{F}= 1$), (ii) emergence of slight differences at $q \approx \pm 4$ ($\mathcal{F} \approx 0.99$), and (iii) a significant difference at  $q\approx \pm 13$ ($\mathcal{F}\approx0.9$). Specifically, the more populated spin component resembles the single-mode approximation (SMA) prediction while the less populated component shows a higher peak at the trap center. 
		
		The analytical result Eq.~(\ref{eqrho_general}) provides explanations for all key features in Fig.~2 of the main text and the corresponding perturbative results in Fig.~\ref{FIG4}. 
		Specifically, for $q=0$, the SPRDMs of the two spin components share the same function form, yielding a unit fidelity ${\mathcal F}=1$. Within the perturbative framework, the decrease of ${\mathcal F}$ with increasing $|q|$ can be directly traced back to the $q$-dependent coefficients of the less populated spin component. As $|q|$ increases, the less populated spin component deviates gradually away from the SMA, while the more populated remains nearly unchanged. Consequently, the SPRDMs of the two spin components diverge​ increasingly, leading directly to the observed reduction in ${\mathcal F}$.
		
		The dependence of $\mathcal F$ on $c_2$  can be understood from the SPRDM structure for each spin component. For the less populated spin component, any dependence on $c_2$ is removed upon normalization, rendering the SPRDM insensitive to both the magnitude and sign of $c_2$. In contrast, the SPRDM of the more populated spin component depends solely on $c_0$ for $q>0$, whereas for $q<0$ it involves the combination $c_0 - c_2$ and introduces a quantitative weak dependence on $c_2 $. This analysis explains the behavior shown in Fig.~\ref{FIG4}(a).
		
		We now turn to the density distribution $n_{m_F}(x)$. For the less populated spin component, the central peak arises from constructive interference between different spatial modes, particularly the terms involving $\varphi_1$ and $\varphi_3$. For the more populated spin component, the central density is dominated by the strong occupation of the ground-state spatial mode $\varphi_1$, effectively satisfying the SMA. In both cases, these mechanisms yield a high density near the trap center that gradually decreases towards the boundaries. In summary, the perturbation analysis in this appendix provides a complete analytical foundation for the phenomena reported in the main text.
		
		\section{Effective three-level model}
		\label{sec:threelevel}
		
		In Sec.~\ref{dynamics}, we observe that at specific values of $\tilde{q} \equiv q/(|c_2| n)$, the dynamics of the spin $m_F=0$ population exhibits a two-timescale structure: a small-amplitude high-frequency oscillation (micromotion) superimposed on a large-amplitude low-frequency modulation (macromotion). Here, we demonstrate that this phenomenon can be quantitatively captured by an effective three-level model derived from the full many-body Hamiltonian Eq.~(\ref{H_Definition}).
		
		We consider the case of $^{7}\mathrm{Li}$ as an example without loss of generality. At a representative resonance point $\tilde{q} =\tilde{q} _{\mathrm{r}}=14.5$, the dynamics is dominated by three basis states constructed as direct products of spatial and spin degrees of freedom. These states are given by
		\begin{align*}
			|1\rangle &= |\varphi_1(x_1)\varphi_1(x_2)\rangle \otimes |0,0\rangle, \\
			|2\rangle &= |\varphi_1(x_1)\varphi_1(x_2)\rangle \otimes \frac{|1,-1\rangle + |-1,1\rangle}{\sqrt{2}}, \\
			|3\rangle &=\frac{|\varphi_{n_i}(x_1)\varphi_{n_j}(x_2)\rangle+|\varphi_{n_j}(x_1)\varphi_{n_i}(x_2)\rangle}{\sqrt{2(1+\delta{n_i,n_j})}} \otimes |0,0\rangle
		\end{align*}
		where $|\varphi_n(x)\rangle$ denotes the single-particle eigenstates of the one-dimensional infinite square well, and the indices $(n_i,n_j)$ label an excited spatial configuration. Projecting the full Hamiltonian of Eq.~(\ref{H_Definition}) onto the subspace spanned by $\{|1\rangle, |2\rangle, |3\rangle\}$ yields an effective three-level Hamiltonian of the form
		\begin{align*}
			H_{\mathrm{eff}} =
			\begin{pmatrix}
				E_1 & V_{12} & V_{13} \\
				V_{12} & E_2 & V_{23} \\
				V_{13} & V_{23} & E_3
			\end{pmatrix}
		\end{align*}
		where $V_{12}$ and $V_{13}$ describe the coupling between the ground state and the excited states. These couplings originate from the spin-independent term with $c_0$ and the spin-dependent interaction with $c_2$. The Zeeman term contributes only to the diagonal energy of spin states with $\left(|1,-1\rangle+|-1,1\rangle\right)/\sqrt{2}$ and does not generate off-diagonal coupling. The initial state is $|\Psi\rangle=a\,|1\rangle +b |2\rangle =\left(\sqrt{2}\,|1\rangle + |2\rangle\right)/\sqrt{3}$, which is consistent with the main text.

		﻿\begin{figure}[tb]
			\centering
			\includegraphics[width=8.6cm]{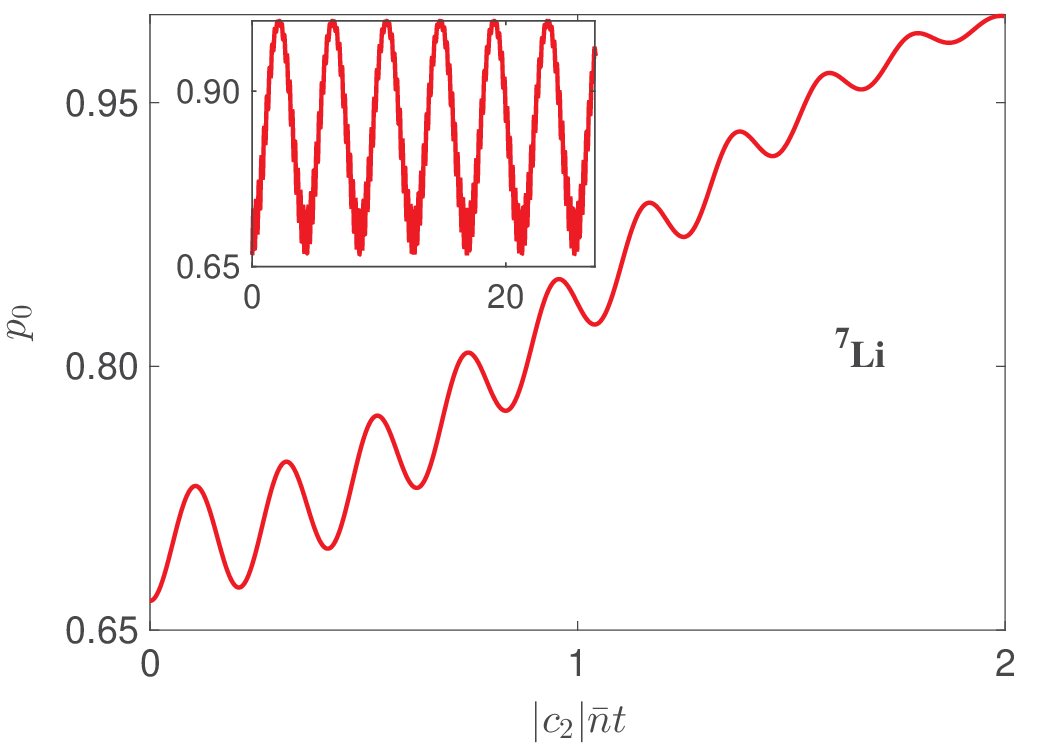}
			\caption{
				\label{FIG5} 
				Time evolution of $p_0(t)$ at a resonant quadratic Zeeman energy $\tilde{q} =\tilde{q} _{\mathrm{r}}=14.5$ for $^{7}\text{Li}$ atoms. The main panel shows the short-time dynamics, and the inset displays a long-time window highlighting the slow oscillation with a large amplitude. The coexistence of these two frequencies directly manifests the effective three-level structure discussed in the text. 
			}
		\end{figure}
		
		Away from the resonant spin-spatial coupling, the observed dynamics is dominated by a single high-frequency component. This behavior is consistent with the SMA prediction, where only these two states \(|1\rangle\) and \(|2\rangle\) are involved, and other states are neglected. The effective Hamiltonian reduces to 
		\begin{equation*}
			H_{\mathrm{SMA}} =
			\begin{pmatrix}
				E_1 & V_{12} \\
				V_{12}   & E_2
			\end{pmatrix}
			\label{eq:H2}
		\end{equation*}
		where
		\[
		E_1 = 2\varepsilon_1 + \frac{3c_0}{2},\,
		E_2 = 2\varepsilon_1 + 2q + \frac{3(c_0 - c_2)}{2}, \,
		V_{12} = \frac{3\sqrt{2}\,c_2}{2}.
		\]
		
		We define the detuning $\Delta = E_2 - E_1$ and the generalized Rabi frequency $\Omega = \sqrt{\Delta^{2} + 4 V_{12}^{\,2}}$. The oscillation frequency can then be written as
		\begin{equation*}
			\Omega_{\mathrm{SMA}} = \sqrt{\Bigl(2q - \frac{3c_{2}}{2}\Bigr)^{2} + 18 c_{2}^{\,2}}\;.
			\label{eq:Omega_SMA}
		\end{equation*}
		The oscillation amplitude is given by
		\begin{equation}
			A = \Biggl| a^{2} - \biggl( a \frac{\Delta}{\Omega_{\mathrm{SMA}}}
			- 2b \frac{V_{12}}{\Omega_{\mathrm{SMA}}} \biggr)^{\!2} \Biggr| .
			\label{eq:A_osc}
		\end{equation}
		where \(a = \sqrt{2/3}\) and \(b = \sqrt{1/3}\) for the given initial state. The analytical expression in Eq.~(\ref{eq:A_osc}) quantitatively reproduces the dependence of the oscillation amplitude on $q$. For positive $q$, the amplitude will reach a maximum value,
		\[
		A_{\mathrm{m}} = 
		\begin{cases}
			\displaystyle \frac{2}{3}, & c_2 > 0,\;\; 	q = \frac{9}{4}c_2,\\[8pt]
			\displaystyle \frac{1}{3}, & c_2 < 0,\;\; q = -\frac{9}{4}c_2 .
		\end{cases}
		\]
		
		Expressed in terms of $\tilde{q}$, the location of the maximum amplitude is therefore $\tilde{q}_{\mathrm{m}} = 9/8$ independent of \(c_2\). This behavior is apparently consistent with the SMA prediction.
		
		As $\tilde{q}$ increases, the state $|2\rangle$ can be tuned into near-degeneracy with the state $|3\rangle$. A resonant enhancement of the spin dynamics occurs when the near-degeneracy condition is satisfied, $E_2 \approx E_3$. Although varying \(\tilde q\) may, in principle, bring \(|2\rangle\) close in energy to many excited states, only those excited states with nonzero interaction induced coupling $V_{23}$ can exhibit resonance dynamics. As a result, observable resonant enhancement of the $p_0(t)$ oscillation amplitude occurs only at certain discrete values of $\tilde q$.
		
		The time evolution of \(p_0(t)\) obtained from the effective three-level model Hamiltonian \(H_{\mathrm{eff}}\) exhibits two well-separated oscillations, as shown in Fig.~\ref{FIG5}, which are qualitatively consistent with the main text. The fast oscillation corresponds to the SMA-like carrier associated with the transition between \(|1\rangle\) and \(|2\rangle\). Near resonance, the coupling \(V_{23}\) hybridizes \(|2\rangle\) and \(|3\rangle\), so this high-frequency component can split into two nearby branches in the Fourier spectrum. Since the two branches are close to each other, they mainly appear in the time-domain signal \(p_0(t)\) as a single fast carrier modulated by the slow macromotion. In contrast, a slow oscillation with a much larger amplitude emerges at long times. This slow dynamics arises from the coherent population exchange between the nearly degenerate states $| 2 \rangle $ and  $| 3 \rangle $. The resonance leads to a slow Rabi-like oscillation with a characteristic frequency $\omega_{\mathrm{slow}} \simeq \sqrt{(E_2 - E_3)^2 + 4|V_{23}|^2}$, which reduces to \(\omega_{\mathrm{slow}} \approx 2|V_{23}|\) on resonance. Since \(V_{23}\) is proportional to $c_2$, the slow oscillation frequency is obviously controlled by the interaction strength. This two-timescale structure is consistent with the main text, exhibiting distinct features beyond SMA dynamics and highlights the key role of other spatial modes in shaping the system's spin dynamics.
		﻿ 
		
	\end{appendix}

	\bibliographystyle{apsrev4-2}
	\bibliography{refs}

@article{Armijo2010ProbingThreeBodyCorrelations,
  title = {Probing {{Three-Body Correlations}} in a {{Quantum Gas Using}} the {{Measurement}} of the {{Third Moment}} of {{Density Fluctuations}}},
  author = {Armijo, J. and Jacqmin, T. and Kheruntsyan, K. V. and Bouchoule, I.},
  year = 2010,
  month = nov,
  journal = {Phys. Rev. Lett.},
  volume = {105},
  number = {23},
  pages = {230402},
  issn = {0031-9007, 1079-7114},
  doi = {10.1103/PhysRevLett.105.230402},
  urldate = {2025-06-24},
  copyright = {http://link.aps.org/licenses/aps-default-license},
  langid = {english}
}

@article{Bravyi2011SchriefferWolffTransformation,
  title = {Schrieffer--{{Wolff}} Transformation for Quantum Many-Body Systems},
  author = {Bravyi, Sergey and DiVincenzo, David P. and Loss, Daniel},
  year = 2011,
  month = oct,
  journal = {Ann. Phys.},
  volume = {326},
  number = {10},
  pages = {2793--2826},
  issn = {00034916},
  doi = {10.1016/j.aop.2011.06.004},
  urldate = {2026-05-13},
  copyright = {https://www.elsevier.com/tdm/userlicense/1.0/},
  langid = {english}
}

@article{Cao2007PairedStateIntegrable,
  title = {Paired State in an Integrable Spin-1 Boson Model},
  author = {Cao, Junpeng and {Yuzhu Jiang} and Wang, Yupeng},
  year = 2007,
  month = aug,
  journal = {EPL},
  volume = {79},
  number = {3},
  pages = {30005},
  issn = {0295-5075, 1286-4854},
  doi = {10.1209/0295-5075/79/30005},
  urldate = {2024-07-26},
  langid = {english}
}

@article{Cao2025JointEstimationTwoPhase,
  title = {Joint {{Estimation}} of a {{Two-Phase Spin Rotation}} beyond {{Classical Limit}}},
  author = {Cao, Jiahao and Li, Xinwei and Mao, Tianwei and Xu, Wenxin and You, Li},
  year = 2025,
  month = jul,
  journal = {Phys. Rev. Lett.},
  volume = {135},
  number = {2},
  pages = {023403},
  issn = {0031-9007, 1079-7114},
  doi = {10.1103/n4f7-7vd1},
  urldate = {2026-01-31},
  langid = {english}
}

@article{Chang2005CoherentSpinorDynamics,
  title = {Coherent Spinor Dynamics in a Spin-1 {{Bose}}~Condensate},
  author = {Chang, Ming-Shien and Qin, Qishu and Zhang, Wenxian and You, Li and Chapman, Michael S.},
  year = 2005,
  month = nov,
  journal = {Nat. Phys.},
  volume = {1},
  number = {2},
  pages = {111--116},
  issn = {1745-2481},
  doi = {10.1038/nphys153}
}

@article{Chomaz2015EmergenceCoherenceTransverse,
  title = {Emergence of Coherence via Transverse Condensation in a Uniform Quasi-Two-Dimensional {{Bose}} Gas},
  author = {Chomaz, Lauriane and Corman, Laura and Bienaim{\'e}, Tom and Desbuquois, R{\'e}mi and Weitenberg, Christof and Nascimb{\`e}ne, Sylvain and Beugnon, J{\'e}r{\^o}me and Dalibard, Jean},
  year = 2015,
  month = jan,
  journal = {Nat. Commun.},
  volume = {6},
  number = {1},
  pages = {6162},
  issn = {2041-1723},
  doi = {10.1038/ncomms7162},
  urldate = {2025-09-18},
  langid = {english}
}

@article{Dag2018ClassificationQuenchdynamicalBehaviors,
  title = {Classification of Quench-Dynamical Behaviors in Spinor Condensates},
  author = {Da{\u g}, Ceren B. and Wang, Sheng-Tao and Duan, L.-M.},
  year = 2018,
  month = feb,
  journal = {Phys. Rev. A},
  volume = {97},
  number = {2},
  pages = {023603},
  issn = {2469-9926, 2469-9934},
  doi = {10.1103/PhysRevA.97.023603},
  urldate = {2025-06-21},
  langid = {english}
}

@article{Devarajan2008GeneralizedVanVleck,
  title = {Generalized van {{Vleck Perturbation Theory}} ({{GVVPT2}}) {{Study}} of the {{Excited States}} of {{Benzene}} and the {{Azabenzenes}}},
  author = {Devarajan, Ajitha and Gaenko, Alexander V. and Khait, Yuri G. and Hoffmann, Mark R.},
  year = 2008,
  month = mar,
  journal = {J. Phys. Chem. A},
  volume = {112},
  number = {12},
  pages = {2677--2682},
  issn = {1089-5639, 1520-5215},
  doi = {10.1021/jp077702w},
  urldate = {2026-05-13},
  langid = {english}
}

@article{Evrard2021ObservationFragmentationSpinor,
  title = {Observation of Fragmentation of a Spinor {{Bose-Einstein}} Condensate},
  author = {Evrard, Bertrand and Qu, An and Dalibard, Jean and Gerbier, Fabrice},
  year = 2021,
  month = sep,
  journal = {Science},
  volume = {373},
  number = {6561},
  pages = {1340--1343},
  issn = {0036-8075, 1095-9203},
  doi = {10.1126/science.abd8206},
  urldate = {2026-01-31},
  langid = {english}
}

@article{Gaudin1971BoundaryEnergyBose,
  title = {Boundary {{Energy}} of a {{Bose Gas}} in {{One Dimension}}},
  author = {Gaudin, M.},
  year = 1971,
  month = jul,
  journal = {Phys. Rev. A},
  volume = {4},
  number = {1},
  pages = {386--394},
  issn = {0556-2791},
  doi = {10.1103/PhysRevA.4.386},
  urldate = {2024-07-26},
  copyright = {http://link.aps.org/licenses/aps-default-license},
  langid = {english}
}

@article{Gaunt2013BoseEinsteinCondensationAtoms,
  title = {Bose-{{Einstein Condensation}} of {{Atoms}} in a {{Uniform Potential}}},
  author = {Gaunt, Alexander L. and Schmidutz, Tobias F. and Gotlibovych, Igor and Smith, Robert P. and Hadzibabic, Zoran},
  year = 2013,
  month = may,
  journal = {Phys. Rev. Lett.},
  volume = {110},
  number = {20},
  pages = {200406},
  issn = {0031-9007, 1079-7114},
  doi = {10.1103/PhysRevLett.110.200406},
  urldate = {2025-09-18},
  copyright = {http://link.aps.org/licenses/aps-default-license},
  langid = {english}
}

@article{Gu2003FerromagneticPhaseTransition,
  title = {Ferromagnetic Phase Transition and {{Bose-Einstein}} Condensation in Spinor {{Bose}} Gases},
  author = {Gu, Qiang and Klemm, Richard A.},
  year = 2003,
  month = sep,
  journal = {Phys. Rev. A},
  volume = {68},
  number = {3},
  pages = {031604},
  issn = {1050-2947, 1094-1622},
  doi = {10.1103/PhysRevA.68.031604},
  urldate = {2025-06-21},
  copyright = {http://link.aps.org/licenses/aps-default-license},
  langid = {english}
}

@article{He2015CoherentSpinmixingDynamics,
  title = {Coherent Spin-Mixing Dynamics in Thermal \$\textasciicircum\textbraceleft 87\textbraceright\textbackslash mathrm\textbraceleft{{Rb}}\textbraceright\$ Spin-1 and Spin-2 Gases},
  author = {He, Xiaodong and Zhu, Bing and Li, Xiaoke and Wang, Fudong and Xu, Zhi-Fang and Wang, Dajun},
  year = 2015,
  month = mar,
  journal = {Phys. Rev. A},
  volume = {91},
  number = {3},
  pages = {033635},
  publisher = {American Physical Society},
  doi = {10.1103/PhysRevA.91.033635},
  urldate = {2026-05-14}
}

@article{Higbie2005DirectNondestructiveImaging,
  title = {Direct {{Nondestructive Imaging}} of {{Magnetization}} in a {{Spin-1 Bose-Einstein Gas}}},
  author = {Higbie, J. M. and Sadler, L. E. and Inouye, S. and Chikkatur, A. P. and Leslie, S. R. and Moore, K. L. and Savalli, V. and {Stamper-Kurn}, D. M.},
  year = 2005,
  month = jul,
  journal = {Phys. Rev. Lett.},
  volume = {95},
  number = {5},
  pages = {050401},
  issn = {0031-9007, 1079-7114},
  doi = {10.1103/PhysRevLett.95.050401},
  urldate = {2026-02-08},
  copyright = {http://link.aps.org/licenses/aps-default-license},
  langid = {english}
}

@article{Ho1998SpinorBoseCondensates,
  title = {Spinor {{Bose Condensates}} in {{Optical Traps}}},
  author = {Ho, Tin-Lun},
  year = 1998,
  month = jul,
  journal = {Phys. Rev. Lett.},
  volume = {81},
  number = {4},
  pages = {742--745},
  doi = {10.1103/PhysRevLett.81.742},
  langid = {english}
}

@article{Huh2020ObservationStronglyFerromagnetic,
  title = {Observation of a Strongly Ferromagnetic Spinor {{Bose-Einstein}} Condensate},
  author = {Huh, SeungJung and Kim, Kyungtae and Kwon, Kiryang and Choi, Jae-yoon},
  year = 2020,
  month = sep,
  journal = {Phys. Rev. Res.},
  volume = {2},
  number = {3},
  pages = {033471},
  issn = {2643-1564},
  doi = {10.1103/PhysRevResearch.2.033471},
  urldate = {2025-10-20},
  langid = {english}
}

@article{Isoshima1999SpindomainFormationSpinor,
  title = {Spin-Domain Formation in Spinor {{Bose-Einstein}} Condensation},
  author = {Isoshima, Tomoya and Machida, Kazushige and Ohmi, Tetsuo},
  year = 1999,
  month = dec,
  journal = {Phys. Rev. A},
  volume = {60},
  number = {6},
  pages = {4857--4863},
  issn = {1050-2947, 1094-1622},
  doi = {10.1103/PhysRevA.60.4857},
  urldate = {2025-12-18},
  copyright = {http://link.aps.org/licenses/aps-default-license},
  langid = {english}
}

@article{Jaksch1998ColdBosonicAtoms,
  title = {Cold {{Bosonic Atoms}} in {{Optical Lattices}}},
  author = {Jaksch, D. and Bruder, C. and Cirac, J. I. and Gardiner, C. W. and Zoller, P.},
  year = 1998,
  month = oct,
  journal = {Phys. Rev. Lett.},
  volume = {81},
  number = {15},
  pages = {3108--3111},
  issn = {0031-9007, 1079-7114},
  doi = {10.1103/PhysRevLett.81.3108},
  urldate = {2025-06-25},
  copyright = {http://link.aps.org/licenses/aps-default-license},
  langid = {english}
}

@article{Jie2020MeanfieldSpinoscillationDynamics,
  title = {Mean-Field Spin-Oscillation Dynamics beyond the Single-Mode Approximation for a Harmonically Trapped Spin-1 {{Bose-Einstein}} Condensate},
  author = {Jie, Jianwen and Guan, Q. and Zhong, S. and Schwettmann, A. and Blume, D.},
  year = 2020,
  month = aug,
  journal = {Phys. Rev. A},
  volume = {102},
  number = {2},
  pages = {023324},
  issn = {2469-9926, 2469-9934},
  doi = {10.1103/PhysRevA.102.023324},
  urldate = {2025-06-27},
  langid = {english}
}

@article{Jie2023DynamicalMeanfielddrivenSpinorcondensate,
  title = {Dynamical Mean-Field-Driven Spinor-Condensate Physics beyond the Single-Mode Approximation},
  author = {Jie, Jianwen and Zhong, Shan and Zhang, Qimin and Morgenstern, Isaiah and Ooi, Hio Giap and Guan, Q. and Bhagat, Anita and Nematollahi, Delaram and Schwettmann, A. and Blume, D.},
  year = 2023,
  month = may,
  journal = {Phys. Rev. A},
  volume = {107},
  number = {5},
  pages = {053309},
  issn = {2469-9926, 2469-9934},
  doi = {10.1103/PhysRevA.107.053309},
  urldate = {2025-06-26},
  langid = {english}
}

@article{JimenezGarcia2019SpontaneousFormationRelaxation,
  title = {Spontaneous Formation and Relaxation of Spin Domains in Antiferromagnetic Spin-1 Condensates},
  author = {Jim{\'e}nez Garc{\'i}a, K. and Invernizzi, A. and Evrard, B. and Frapolli, C. and Dalibard, J. and Gerbier, F.},
  year = 2019,
  month = mar,
  journal = {Nat. Commun.},
  volume = {10},
  number = {1},
  pages = {1422},
  issn = {2041-1723},
  doi = {10.1038/s41467-019-08505-6},
  urldate = {2025-11-27},
  langid = {english}
}

@article{Jordahl1934EffectCrystallineElectric,
  title = {The {{Effect}} of {{Crystalline Electric Fields}} on the {{Paramagnetic Susceptibility}} of {{Cupric Salts}}},
  author = {Jordahl, O. M.},
  year = 1934,
  month = jan,
  journal = {Phys. Rev.},
  volume = {45},
  number = {2},
  pages = {87--97},
  issn = {0031-899X},
  doi = {10.1103/PhysRev.45.87},
  urldate = {2025-04-23},
  copyright = {http://link.aps.org/licenses/aps-default-license},
  langid = {english}
}

@article{Jozsa1994FidelityMixedQuantum,
  title = {Fidelity for {{Mixed Quantum States}}},
  author = {Jozsa, Richard},
  year = 1994,
  month = dec,
  journal = {J. Mod. Opt.},
  volume = {41},
  number = {12},
  pages = {2315--2323},
  issn = {0950-0340, 1362-3044},
  doi = {10.1080/09500349414552171},
  urldate = {2025-11-06},
  langid = {english}
}

@article{Kawaguchi2012SpinorBoseEinsteina,
  title = {Spinor Bose--Einstein Condensates},
  author = {Kawaguchi, Yuki and Ueda, Masahito},
  year = 2012,
  journal = {Phys. Rep.},
  volume = {520},
  number = {5},
  pages = {253--381},
  issn = {0370-1573},
  doi = {10.1016/j.physrep.2012.07.005},
  langid = {english}
}

@article{Kim2021EmissionSpinCorrelatedMatterWavea,
  title = {Emission of {{Spin-Correlated Matter-Wave Jets}} from {{Spinor Bose-Einstein Condensates}}},
  author = {Kim, Kyungtae and Hur, Junhyeok and Huh, SeungJung and Choi, Soonwon and Choi, Jae-yoon},
  year = 2021,
  month = jul,
  journal = {Phys. Rev. Lett.},
  volume = {127},
  number = {4},
  pages = {043401},
  issn = {0031-9007, 1079-7114},
  doi = {10.1103/PhysRevLett.127.043401},
  urldate = {2026-01-31},
  langid = {english}
}

@article{Kirtman1981SimultaneousCalculationSeveral,
  title = {Simultaneous Calculation of Several Interacting Electronic States by Generalized {{Van Vleck}} Perturbation Theory},
  author = {Kirtman, Bernard},
  year = 1981,
  month = jul,
  journal = {J. Chem. Phys.},
  volume = {75},
  number = {2},
  pages = {798--808},
  issn = {0021-9606, 1089-7690},
  doi = {10.1063/1.442123},
  urldate = {2026-06-01},
  langid = {english}
}

@article{Lan2014RamandressedSpin1Spinorbitcoupled,
  title = {Raman-Dressed Spin-1 Spin-Orbit-Coupled Quantum Gas},
  author = {Lan, Zhihao and {\"O}hberg, Patrik},
  year = 2014,
  month = feb,
  journal = {Phys. Rev. A},
  volume = {89},
  number = {2},
  pages = {023630},
  issn = {1050-2947, 1094-1622},
  doi = {10.1103/PhysRevA.89.023630},
  urldate = {2025-06-21},
  copyright = {http://link.aps.org/licenses/aps-default-license},
  langid = {english}
}

@article{Law1998QuantumSpinsMixing,
  title = {Quantum {{Spins Mixing}} in {{Spinor Bose-Einstein Condensates}}},
  author = {Law, C K and Pu, H and Bigelow, N P},
  year = 1998,
  journal = {Phys. Rev. Lett.},
  volume = {81},
  number = {24},
  pages = {5257--5261},
  doi = {10.1103/PhysRevLett.81.5257},
  langid = {english}
}

@article{Leibfried2003QuantumDynamicsSinglea,
  title = {Quantum Dynamics of Single Trapped Ions},
  author = {Leibfried, D. and Blatt, R. and Monroe, C. and Wineland, D.},
  year = 2003,
  month = mar,
  journal = {Rev. Mod. Phys.},
  volume = {75},
  number = {1},
  pages = {281--324},
  issn = {0034-6861, 1539-0756},
  doi = {10.1103/RevModPhys.75.281},
  urldate = {2026-01-31},
  copyright = {http://link.aps.org/licenses/aps-default-license},
  langid = {english}
}

@article{Li2020ManipulationHeteronuclearSpin,
  title = {Manipulation of Heteronuclear Spin Dynamics with Microwave and Vector Light Shift},
  author = {Li, Lintao and Zhu, Bing and Lu, Bo and Zhang, Shizhong and Wang, Dajun},
  year = 2020,
  month = may,
  journal = {Phys. Rev. A},
  volume = {101},
  number = {5},
  pages = {053611},
  issn = {2469-9926, 2469-9934},
  doi = {10.1103/PhysRevA.101.053611},
  urldate = {2026-05-15},
  langid = {english}
}

@article{Liu2022NonlinearInterferometryClassical,
  title = {Nonlinear Interferometry beyond Classical Limit Enabled by Cyclic Dynamics},
  author = {Liu, Qi and Wu, Ling-Na and Cao, Jia-Hao and Mao, Tian-Wei and Li, Xin-Wei and Guo, Shuai-Feng and Tey, Meng Khoon and You, Li},
  year = 2022,
  month = feb,
  journal = {Nat. Phys.},
  volume = {18},
  number = {2},
  pages = {167--171},
  issn = {1745-2473, 1745-2481},
  doi = {10.1038/s41567-021-01441-7},
  urldate = {2026-01-31},
  langid = {english}
}

@article{Matuszewski2009ExcitedSpinStates,
  title = {Excited Spin States and Phase Separation in Spinor {{Bose-Einstein}} Condensates},
  author = {Matuszewski, Micha{\l} and Alexander, Tristram J. and Kivshar, Yuri S.},
  year = 2009,
  month = aug,
  journal = {Phys. Rev. A},
  volume = {80},
  number = {2},
  pages = {023602},
  issn = {1050-2947, 1094-1622},
  doi = {10.1103/PhysRevA.80.023602},
  urldate = {2025-11-27},
  copyright = {http://link.aps.org/licenses/aps-default-license},
  langid = {english}
}

@article{Mestrom2021ThreebodySpinMixing,
  title = {Three-Body Spin Mixing in Spin-1 {{Bose-Einstein}} Condensates},
  author = {Mestrom, P. M. A. and Li, J.-L. and Colussi, V. E. and Secker, T. and Kokkelmans, S. J. J. M. F.},
  year = 2021,
  month = aug,
  journal = {Phys. Rev. A},
  volume = {104},
  number = {2},
  pages = {023321},
  issn = {2469-9926, 2469-9934},
  doi = {10.1103/PhysRevA.104.023321},
  urldate = {2026-05-15},
  langid = {english}
}

@article{Mistakidis2023FewbodyBoseGases,
  title = {Few-Body {{Bose}} Gases in Low Dimensions---{{A}} Laboratory for Quantum Dynamics},
  author = {Mistakidis, S.I. and Volosniev, A.G. and Barfknecht, R.E. and Fogarty, T. and Busch, {\relax Th}. and Foerster, A. and Schmelcher, P. and Zinner, N.T.},
  year = 2023,
  month = nov,
  journal = {Phys. Rep.},
  volume = {1042},
  pages = {1--108},
  issn = {03701573},
  doi = {10.1016/j.physrep.2023.10.004},
  urldate = {2025-03-31},
  langid = {english}
}

@book{Nielsen2010QuantumComputationQuantum,
  title = {Quantum Computation and Quantum Information},
  author = {Nielsen, Michael A. and Chuang, Isaac L.},
  year = 2010,
  publisher = {Cambridge University Press},
  address = {Cambridge},
  isbn = {978-1-107-00217-3},
  langid = {english},
  lccn = {530.12}
}

@article{Ohmi1998BoseEinsteinCondensationInternal,
  title = {Bose-{{Einstein Condensation}} with {{Internal Degrees}} of {{Freedom}} in {{Alkali Atom Gases}}},
  author = {Ohmi, Tetsuo and Machida, Kazushige},
  year = 1998,
  month = jun,
  journal = {J. Phys. Soc. Jpn.},
  volume = {67},
  number = {6},
  pages = {1822--1825},
  issn = {0031-9015, 1347-4073},
  doi = {10.1143/JPSJ.67.1822},
  urldate = {2025-10-20},
  langid = {english}
}

@article{Paredes2004TonksGirardeauGas,
  title = {Tonks--{{Girardeau}} Gas of Ultracold Atoms in an Optical Lattice},
  author = {Paredes, Bel{\'e}n and Widera, Artur and Murg, Valentin and Mandel, Olaf and F{\"o}lling, Simon and Cirac, Ignacio and Shlyapnikov, Gora V. and H{\"a}nsch, Theodor W. and Bloch, Immanuel},
  year = 2004,
  month = may,
  journal = {Nature},
  volume = {429},
  number = {6989},
  pages = {277--281},
  issn = {0028-0836, 1476-4687},
  doi = {10.1038/nature02530},
  urldate = {2025-06-24},
  copyright = {http://www.springer.com/tdm},
  langid = {english}
}

@article{Paul1990ElectromagneticTrapsChargedb,
  title = {Electromagnetic Traps for Charged and Neutral Particles},
  author = {Paul, Wolfgang},
  year = 1990,
  month = jul,
  journal = {Rev. Mod. Phys.},
  volume = {62},
  number = {3},
  pages = {531--540},
  issn = {0034-6861, 1539-0756},
  doi = {10.1103/RevModPhys.62.531},
  urldate = {2026-01-31},
  copyright = {http://link.aps.org/licenses/aps-default-license},
  langid = {english}
}

@article{Pechkis2013SpinorDynamicsAntiferromagnetic,
  title = {Spinor {{Dynamics}} in an {{Antiferromagnetic Spin-1 Thermal Bose Gas}}},
  author = {Pechkis, H. K. and Wrubel, J. P. and Schwettmann, A. and Griffin, P. F. and Barnett, R. and Tiesinga, E. and Lett, P. D.},
  year = 2013,
  month = jul,
  journal = {Phys. Rev. Lett.},
  volume = {111},
  number = {2},
  pages = {025301},
  issn = {0031-9007, 1079-7114},
  doi = {10.1103/PhysRevLett.111.025301},
  urldate = {2026-05-14},
  copyright = {http://link.aps.org/licenses/aps-default-license},
  langid = {english}
}

@article{Pu1999SpinmixingDynamicsSpinor,
  title = {Spin-Mixing Dynamics of a Spinor {{Bose-Einstein}} Condensate},
  author = {Pu, H and Law, C K and Raghavan, S and Eberly, J H and Bigelow, N P},
  year = 1999,
  month = aug,
  journal = {Phys. Rev. A},
  volume = {60},
  number = {2},
  pages = {1463--1470},
  doi = {10.1103/PhysRevA.60.1463},
  langid = {english}
}

@article{Redmon1982MultidimensionalManybodyTheory,
  title = {Multidimensional Many-Body Theory: {{Diagrammatic}} Implementation of a Canonical van {{Vleck}} Formalism},
  shorttitle = {Multidimensional Many-Body Theory},
  author = {Redmon, Lynn T. and Bartlett, Rodney J.},
  year = 1982,
  month = feb,
  journal = {J. Chem. Phys.},
  volume = {76},
  number = {4},
  pages = {1938--1948},
  issn = {0021-9606, 1089-7690},
  doi = {10.1063/1.443167},
  urldate = {2026-06-01},
  langid = {english}
}

@article{Stamper-Kurn1998OpticalConfinementBoseEinstein,
  title = {Optical {{Confinement}} of a {{Bose-Einstein Condensate}}},
  author = {{Stamper-Kurn}, D. M. and Andrews, M. R. and Chikkatur, A. P. and Inouye, S. and Miesner, H.-J. and Stenger, J. and Ketterle, W.},
  year = 1998,
  month = mar,
  journal = {Phys. Rev. Lett.},
  volume = {80},
  number = {10},
  pages = {2027--2030},
  issn = {0031-9007, 1079-7114},
  doi = {10.1103/PhysRevLett.80.2027},
  urldate = {2025-04-23},
  copyright = {http://link.aps.org/licenses/aps-default-license},
  langid = {english}
}

@article{Stamper-Kurn2013SpinorBoseGases,
  title = {Spinor {{Bose}} Gases: {{Symmetries}}, Magnetism, and Quantum Dynamics},
  shorttitle = {Spinor {{Bose}} Gases},
  author = {{Stamper-Kurn}, Dan M. and Ueda, Masahito},
  year = 2013,
  month = jul,
  journal = {Rev. Mod. Phys.},
  volume = {85},
  number = {3},
  pages = {1191--1244},
  issn = {0034-6861, 1539-0756},
  doi = {10.1103/RevModPhys.85.1191},
  urldate = {2025-04-23},
  copyright = {http://link.aps.org/licenses/aps-default-license},
  langid = {english}
}

@article{Stenger1998SpinDomainsGroundstate,
  title = {Spin Domains in Ground-State {{Bose}}--{{Einstein}} Condensates},
  author = {Stenger, J. and Inouye, S. and {Stamper-Kurn}, D. M. and Miesner, H.-J. and Chikkatur, A. P. and Ketterle, W.},
  year = 1998,
  month = nov,
  journal = {Nature},
  volume = {396},
  number = {6709},
  pages = {345--348},
  issn = {0028-0836, 1476-4687},
  doi = {10.1038/24567},
  urldate = {2025-09-18},
  copyright = {http://www.springer.com/tdm},
  langid = {english}
}

@article{Tajik2019DesigningArbitraryOnedimensional,
  title = {Designing Arbitrary One-Dimensional Potentials on an Atom Chip},
  author = {Tajik, Mohammadamin and Rauer, Bernhard and Schweigler, Thomas and Cataldini, Federica and Sabino, Jo{\~a}o and M{\o}ller, Frederik S. and Ji, Si-Cong and Mazets, Igor E. and Schmiedmayer, J{\"o}rg},
  year = 2019,
  month = nov,
  journal = {Opt. Express},
  volume = {27},
  number = {23},
  pages = {33474},
  issn = {1094-4087},
  doi = {10.1364/OE.27.033474},
  urldate = {2025-09-18},
  langid = {english}
}

@article{VanEs2010BoxTrapsAtom,
  title = {Box Traps on an Atom Chip for One-Dimensional Quantum Gases},
  author = {Van Es, J J P and Wicke, P and Van Amerongen, A H and R{\'e}tif, C and Whitlock, S and Van Druten, N J},
  year = 2010,
  month = aug,
  journal = {J. Phys. B: At. Mol. Opt. Phys.},
  volume = {43},
  number = {15},
  pages = {155002},
  issn = {0953-4075, 1361-6455},
  doi = {10.1088/0953-4075/43/15/155002},
  urldate = {2025-09-18},
  langid = {english}
}

@article{VanVleck1929TypeDoublingElectron,
  title = {On {$\sigma$} -{{Type Doubling}} and {{Electron Spin}} in the {{Spectra}} of {{Diatomic Molecules}}},
  author = {Van Vleck, J. H.},
  year = 1929,
  month = apr,
  journal = {Phys. Rev.},
  volume = {33},
  number = {4},
  pages = {467--506},
  issn = {0031-899X},
  doi = {10.1103/PhysRev.33.467},
  urldate = {2025-04-23},
  copyright = {http://link.aps.org/licenses/aps-default-license},
  langid = {english}
}

@article{Vinit2013AntiferromagneticSpatialOrdering,
  title = {Antiferromagnetic {{Spatial Ordering}} in a {{Quenched One-Dimensional Spinor Gas}}},
  author = {Vinit, A. and Bookjans, E. M. and S{\'a} De Melo, C. A. R. and Raman, C.},
  year = 2013,
  month = apr,
  journal = {Phys. Rev. Lett.},
  volume = {110},
  number = {16},
  pages = {165301},
  issn = {0031-9007, 1079-7114},
  doi = {10.1103/PhysRevLett.110.165301},
  urldate = {2025-06-21},
  copyright = {http://link.aps.org/licenses/aps-default-license},
  langid = {english}
}

@article{Wang2010SpinOrbitCoupledSpinor,
  title = {Spin-{{Orbit Coupled Spinor Bose-Einstein Condensates}}},
  author = {Wang, Chunji and Gao, Chao and Jian, Chao-Ming and Zhai, Hui},
  year = 2010,
  month = oct,
  journal = {Phys. Rev. Lett.},
  volume = {105},
  number = {16},
  pages = {160403},
  issn = {0031-9007, 1079-7114},
  doi = {10.1103/PhysRevLett.105.160403},
  urldate = {2025-06-21},
  copyright = {http://link.aps.org/licenses/aps-default-license},
  langid = {english}
}

@article{Widera2005CoherentCollisionalSpin,
  title = {Coherent {{Collisional Spin Dynamics}} in {{Optical Lattices}}},
  author = {Widera, Artur and Gerbier, Fabrice and F{\"o}lling, Simon and Gericke, Tatjana and Mandel, Olaf and Bloch, Immanuel},
  year = 2005,
  month = nov,
  journal = {Phys. Rev. Lett.},
  volume = {95},
  number = {19},
  pages = {190405},
  issn = {0031-9007, 1079-7114},
  doi = {10.1103/PhysRevLett.95.190405},
  urldate = {2024-07-26},
  copyright = {http://link.aps.org/licenses/aps-default-license},
  langid = {english}
}

@article{Yi2001TrappedCondensatesAtoms,
  title = {Trapped Condensates of Atoms with Dipole Interactions},
  author = {Yi, S. and You, L.},
  year = 2001,
  month = apr,
  journal = {Phys. Rev. A},
  volume = {63},
  number = {5},
  pages = {053607},
  issn = {1050-2947, 1094-1622},
  doi = {10.1103/PhysRevA.63.053607},
  urldate = {2026-01-07},
  copyright = {http://link.aps.org/licenses/aps-default-license},
  langid = {english}
}

@article{Yi2002SinglemodeApproximationSpinor1,
  title = {Single-Mode Approximation in a Spinor-1 Atomic Condensate},
  author = {Yi, S. and M{\"u}stecapl{\i}o{\u g}lu, {\"O}. E. and Sun, C. P. and You, L.},
  year = 2002,
  month = jul,
  journal = {Phys. Rev. A},
  volume = {66},
  number = {1},
  pages = {011601},
  issn = {1050-2947, 1094-1622},
  doi = {10.1103/PhysRevA.66.011601},
  urldate = {2025-06-27},
  copyright = {http://link.aps.org/licenses/aps-default-license},
  langid = {english}
}

@article{Zhang2003MeanFieldGround,
  title = {Mean Field Ground State of a Spin-1 Condensate in a Magnetic Field},
  author = {Zhang, Wenxian and Yi, Su and You, Li},
  year = 2003,
  month = jun,
  journal = {New J. Phys.},
  volume = {5},
  pages = {77--77},
  issn = {1367-2630},
  doi = {10.1088/1367-2630/5/1/377},
  urldate = {2026-01-07},
  langid = {english}
}

@article{Zhang2005CoherentSpinMixing,
  title = {Coherent Spin Mixing Dynamics in a Spin-1 Atomic Condensate},
  author = {Zhang, Wenxian and Zhou, D. L. and Chang, M.-S. and Chapman, M. S. and You, L.},
  year = 2005,
  month = jul,
  journal = {Phys. Rev. A},
  volume = {72},
  number = {1},
  pages = {013602},
  issn = {1050-2947, 1094-1622},
  doi = {10.1103/PhysRevA.72.013602},
  urldate = {2026-02-06},
  copyright = {http://link.aps.org/licenses/aps-default-license},
  langid = {english}
}

@article{Zhang2005DynamicalInstabilityDomain,
  title = {Dynamical {{Instability}} and {{Domain Formation}} in a {{Spin-1 Bose-Einstein Condensate}}},
  author = {Zhang, Wenxian and Zhou, D. L. and Chang, M.-S. and Chapman, M. S. and You, L.},
  year = 2005,
  month = oct,
  journal = {Phys. Rev. Lett.},
  volume = {95},
  number = {18},
  pages = {180403},
  issn = {0031-9007, 1079-7114},
  doi = {10.1103/PhysRevLett.95.180403},
  urldate = {2026-01-07},
  copyright = {http://link.aps.org/licenses/aps-default-license},
  langid = {english}
}

@article{Zhao2014DynamicsSpinorCondensates,
  title = {Dynamics in Spinor Condensates Tuned by a Microwave Dressing Field},
  author = {Zhao, L. and Jiang, J. and Tang, T. and Webb, M. and Liu, Y.},
  year = 2014,
  month = feb,
  journal = {Phys. Rev. A},
  volume = {89},
  number = {2},
  pages = {023608},
  issn = {1050-2947, 1094-1622},
  doi = {10.1103/PhysRevA.89.023608},
  urldate = {2025-04-23},
  copyright = {http://link.aps.org/licenses/aps-default-license},
  langid = {english}
}

@article{Zhao2015AntiferromagneticSpinorCondensates,
  title = {Antiferromagnetic {{Spinor Condensates}} in a {{Two-Dimensional Optical Lattice}}},
  author = {Zhao, L. and Jiang, J. and Tang, T. and Webb, M. and Liu, Y.},
  year = 2015,
  month = jun,
  journal = {Phys. Rev. Lett.},
  volume = {114},
  number = {22},
  pages = {225302},
  issn = {0031-9007, 1079-7114},
  doi = {10.1103/PhysRevLett.114.225302},
  urldate = {2025-06-21},
  copyright = {http://link.aps.org/licenses/aps-default-license},
  langid = {english}
}

@article{Zhou2010SpinDynamicsDomain,
  title = {Spin Dynamics and Domain Formation of a Spinor {{Bose-Einstein}} Condensate in an Optical Cavity},
  author = {Zhou, Lu and Pu, Han and Ling, Hong Y. and Zhang, Keye and Zhang, Weiping},
  year = 2010,
  month = jun,
  journal = {Phys. Rev. A},
  volume = {81},
  number = {6},
  pages = {063641},
  issn = {1050-2947, 1094-1622},
  doi = {10.1103/PhysRevA.81.063641},
  urldate = {2025-06-21},
  copyright = {http://link.aps.org/licenses/aps-default-license},
  langid = {english}
}

\end{document}